\pdfoutput=1
\documentclass[%
 aip,
 amsmath,amssymb,
 reprint,%
]{revtex4-1}

\usepackage[normalem]{ulem}
\usepackage{graphicx}
\usepackage{dcolumn}
\usepackage{bm}

\usepackage[utf8]{inputenc}
\usepackage[T1]{fontenc}
\usepackage{mathptmx}
\usepackage{etoolbox}
\usepackage{xcolor}

\makeatletter
\def\@email#1#2{%
 \endgroup
 \patchcmd{\titleblock@produce}
  {\frontmatter@RRAPformat}
  {\frontmatter@RRAPformat{\produce@RRAP{*#1\href{mailto:#2}{#2}}}\frontmatter@RRAPformat}
  {}{}
}%
\makeatother
\begin{document}

\preprint{AIP/123-QED}

\title[Title]{A modified nudged elastic band algorithm with adaptive spring lengths}
\author{D. Mandelli}
\author{M. Parrinello*}%
 \email{michele.parrinello@iit.it.}
\affiliation{ 
Atomistic Simulations, Italian Institute of Technology, Via Morego, 30 16163 Genova, Italy.
}%

\date{\today}

\begin{abstract}
We present a modified version of the nudged elastic band (NEB) algorithm to find minimum energy paths connecting two known configurations. We show that replacing the harmonic band-energy term with a discretized version of the Onsager-Machlup action leads to a NEB algorithm with adaptive spring lengths that automatically increase the resolution of the minimum energy path around the saddle point of the potential energy surface. The method has the same computational cost per optimization step of the standard NEB algorithm and does not introduce additional parameters. We present applications to the isomerization of alanine dipeptide, the elimination of hydrogen from ethane and the healing of a 5-77-5 defect in graphene.
\end{abstract}

\maketitle

\section{\label{sec:Intro} Introduction}

An important problem in chemistry and materials science is the computation of kinetic rates for transitions between metastable states. Typical examples are chemical reactions, molecular conformational changes and diffusion events in solids. All these transformations are thermally activated processes that occur via a concerted rearrangement of atoms that bring the system from one state to another. During the transformation, the system has to pass through high-energy configurations that represent the kinetic bottleneck of the reaction. Although these phenomena can be often described using classical mechanics, a direct simulation of the dynamics of reaction events is usually impossible because in the presence of energy barriers larger than $k_B T$ unfeasibly long simulation times are needed to observe the transitions of interest. Several advanced algorithms based on molecular dynamics have been developed to compute kinetic rates \cite{Dellago1998a,VanErp2003,Faradjian2004,Tiwary2013,Debnath2020,Mandelli2020}, however, they can be computationally expensive, especially when the electronic degrees of freedom play an important role and quantum mechanical approaches are needed to compute the atomic forces. In all the cases where these direct methods become impractical, transition state theory \cite{Eyring1935,Wigner1938,Keck2007} (TST) provides a cheaper route to compute accurate estimates of transition rates. TST is a purely statistical approach where rates are expressed in terms of quantities that are available directly from the potential energy surface (PES).

When applying TST, the most challenging step is the determination of the relevant saddle points in the multidimensional potential energy landscape. Among the algorithms that have been designed to solve this problem one can distinguish between local methods, like the dimer method \cite{Henkelman1999} and the activation-relaxation-technique \cite{Jay2020}, and the so-called chain-of-states methods \cite{Jonsson1998,Peters2004,E2010}. The first class of algorithms uses only local information of the PES to follow the energy landscape uphill to the saddle point and then downhill to a new minimum. These methods are particularly useful when the final state is not known. On the other hand, when both the initial and final states are known, it is generally more effective to adopt chain-of-states approaches. Here, we focus on the popular nudged elastic band (NEB) algorithm \cite{Henkelman2000,Henkelman2000a}, which is an efficient method to find the minimum energy path (MEP) connecting two endpoint configurations. By definition, the force acting on the atoms is parallel to the MEP, while the energy is stationary along any direction perpendicular to it. In the continuum limit, this is equivalent to the requirement
\begin{equation}
\label{eq1}
    \nabla E({\bf R}_{\lambda})|_{\perp}=0
\end{equation}
where E is the potential energy, $R_{\lambda}$ is a path in configuration space parameterized by $\lambda \in [0,1]$, with fixed endpoints ${\bf R}_{0}$ and ${\bf R}_1$, and only the component of the gradient perpendicular to the path is considered. It follows from equation \eqref{eq1} that the PES saddle points correspond to the maxima of the potential energy along the MEP.

In the NEB algorithm, an initial guess of the path connecting the two endpoints is discretized as a sequence of $N$ images (or replicas) of the system. The images are optimized in a concerted way so as to obtain a path that satisfies equation \eqref{eq1}. The configuration corresponding to the maximum of the potential energy along the MEP identifies the saddle point of interest. Generally, one would like to have a higher density of images in the region of the saddle point since this can improve the accuracy with which the saddle point is obtained by improving the resolution of the discretized MEP. However, the standard NEB algorithm generates replicas that are equally spaced along the path \cite{Jonsson1998,Henkelman2000}. A straightforward way to improve the sampling near the saddle is to increase the overall number of images. Still, this approach would lead to most of the computational time being wasted in optimizing images in irrelevant regions far from the saddle. A more efficient solution is provided by algorithms that keep the number of images as small as possible by focusing on increasing the resolution only around the saddle \cite{Henkelman2000a,Maragakis2002,Kolsbjerg2016}. This is particularly important in expensive quantum mechanical calculations, where the number of replicas is typically limited to the range $5<N<20$.

In this work, we show how a relatively simple modification of the inter-replica harmonic forces used in the standard NEB algorithm automatically increases the MEP resolution   around the saddle point, without the need to increase the number of images. The method takes inspiration from a discretized version of the Onsager-Machlup action \cite{Onsager1953}, which leads to a modified NEB algorithm with adaptive natural spring lengths that become automatically shorter around the stationary points of the PES. We present applications to the isomerization of alanine dipeptide, the elimination of hydrogen from ethane and the healing of a 5-77-5 defect in graphene. In all these cases, the modified algorithm reproduces the MEPs and saddle points obtained via standard NEB calculations while improving the resolution around the saddle point.

The article is organized as follows. In Section \ref{subsec:NEB}, we review the standard NEB algorithm. In Section \ref{subsec:NOM} we introduce our modification. In Section \ref{subsec:Models}, we describe the model systems and the simulation protocols used to test the method. In Section \ref{sec:Results}, we present the results of simulations. In Section \ref{sec:Conclusions}, we present our final remarks.

\section{\label{sec:Methods}Materials and Methods}
\subsection{\label{subsec:NEB}The nudged elastic band algorithm}
When discretized into $N$ steps, a path is described by the ordered set $\{{\bf R}_1,{\bf R}_2,\dots,{\bf R}_N\}$ of replicas of the system, where we denote with ${\bf R}_i$ $i=1,\dots,N$, the system atomic coordinates of the different images. The initial and final positions ${\bf R}_1$ and ${\bf R}_N$ are set at the local minima of the potential energy. The NEB algorithm optimizes the $N-2$ intermediate configurations by putting to zero the forces
\begin{equation}
    \label{eq2}
    {\bf F}_i=-\nabla E({\bf R}_i)|_{\perp}+{\bf F}_i^s|_{||}+{\bf F}_i^{s'}|_{\perp}
\end{equation}
acting on each image. The first term
\begin{equation}
    \label{eq3}
    \nabla E({\bf R}_i)|_{\perp}=\nabla E({\bf R}_i)-\nabla E({\bf R}_i)\cdot{\bf \tau}_i{\bf \tau}_i,
\end{equation}
is the component of the physical force acting perpendicular to the path at image $i$. This term is a discretized version of equation \eqref{eq1} and it ensures convergence to the MEP. In equation \eqref{eq3}, ${\bf \tau}_i$ is the unit vector tangent to the path at image $i$ and ${\bf \tau}_i{\bf \tau}_i$ indicates the outer product. In all our calculations, ${\bf \tau}_i$ was computed using the definition of Ref.\onlinecite{Henkelman2000}. The second term in equation \eqref{eq2} is a fictitious spring force
\begin{equation}
    \label{eq4}
    {\bf F}_i^s|_{||}=k_{||}(|{\bf R}_i-{\bf R}_{i-1}|-|{\bf R}_{i+1}-{\bf R}_i|){\bf \tau}_i,
\end{equation}
acting parallel to the path. This elastic constraint is necessary when dealing with discretized paths in order to prevent the images from sliding downhill and to keep them equally spaced. The spring constant $k_{||}$ is a user-defined parameter. Because only the component parallel to the path is used, typically the results are weakly dependent on the choice of $k_{||}$. The last term is an additional harmonic-like force given by
\begin{equation}
    \label{eq5}
    {\bf F}_i^{s'}|_{\perp}=f(\phi_i)({\bf F}_i^{s'}-{\bf F}_i^{s'}\cdot{\bf \tau}_i{\bf \tau}_i),
\end{equation}
where
\begin{equation}
    \label{eq6}
    {\bf F}_i^{s'}=k_{\perp}({\bf R}_{i+1}+{\bf R}_{i-1}-2{\bf R}_{i}),
\end{equation}
and
\begin{equation}
    f(\phi_i)=\begin{cases}
      \frac{1}{2} (1+\cos(\pi\cos(\phi_i))) & \text{ if $0<\phi_i<\pi/2$}\\
      1 & \text{if $\phi_i>\pi/2$}
    \end{cases}
\end{equation}
is a switching function that depends on the path angle at image $i$ \cite{Jonsson1998,Maras2016}:
\begin{equation}
    \cos\left(\phi_i\right)=\frac{({\bf R}_{i+1} - {\bf R}_i) \cdot ({\bf R}_{i} - {\bf R}_{i-1})}{|{\bf R}_{i+1} - {\bf R}_i||{\bf R}_{i} - {\bf R}_{i-1}|}
\end{equation}

We note that the tangent introduced in Ref.\onlinecite{Henkelman2000} was specifically designed to prevent oscillations of the path without the need of including the perpendicular spring force of equation \eqref{eq5}. However, as recently observed in Ref.\onlinecite{Maras2016}, keeping part of the perpendicular spring force can help improving the convergence. In this work, we have found it useful in some cases to include this term for few initial steps of the minimization. In all cases, to avoid systematic corner cutting problems \cite{Jonsson1998}, the final path was eventually obtained switching off this term by setting $k_{\perp}=0$.

Once the condition ${\bf F}_i=0$ is met, the sequence of images lie along the MEP. However, usually no image relaxes exactly on top of the saddle point. Hence, after a first relaxation, a typical NEB calculation continues with a second “climbing image” optimization \cite{Henkelman2000a} during which the force acting on the replica corresponding to the maximum value of the potential energy is changed to
\begin{equation}
    \label{eq7}
    {\bf F}_i^{\rm climber}=-\nabla E({\bf R}_i)+2\nabla E({\bf R}_i)\cdot{\bf \tau}_i{\bf \tau}_i.
\end{equation}
This is the physical force with the component parallel to the path changed in sign. The effect of this force is to push the climber uphill and towards the saddle point. The climbing image is subjected only to the force defined in equation \eqref{eq7} and it does not feel any spring force, either parallel or perpendicular. Hence, provided the estimate of the tangent to the MEP (${\bf \tau}_i$) is accurate enough, the climbing image will converge to the exact saddle.

\subsection{\label{subsec:NOM}Modified nudged elastic band algorithm}
Equation \eqref{eq4} can be interpreted as the force arising from a fictitious harmonic interaction between the replicas
\begin{equation}
    \label{eq8}
    S_{\rm NEB}=\sum_{i=1}^{N-1}\frac{k_{||}}{2}|{\bf R}_{i+1}-{\bf R}_i|^2
\end{equation}
that keeps the images equally spaced along the path. In order to increase the resolution of the MEP around the saddle point some variants of the NEB algorithm have been proposed that consist in increasing automatically the spring constant $k_{||}$ near the saddle point \cite{Henkelman2000a,Asgeirsson2021} or in adding more images only in its proximity \cite{Maragakis2002,Kolsbjerg2016}.

Here, we propose an alternative approach that exploits the fact that saddles are stationary points of the PES. In designing this method, we were inspired by the Onsager-Machlup (OM) action \cite{Onsager1953}, $S_{\rm OM}[{\bf R}(t)]$, that defines the probability, $p\propto e^{-S_{\rm OM}/k_B T}$, of observing a stochastic trajectory ${\bf R}(t)$. In the case of a discretized Brownian trajectory of $N$ steps, ${\bf R}_1\rightarrow{\bf R}_2\rightarrow\dots\rightarrow{\bf R}_N$, connecting two endpoints ${\bf R}_1$ and ${\bf R}_N$, the OM action is given by \cite{Mandelli2020}
\begin{equation}
    \label{eq9}
    S_{\rm OM}=\sum_{i=1}^{N-1}\frac{m\nu}{4\Delta t}({\bf R}_{i+1}-{\bf R}_i-{\bf L}_i)^2.
\end{equation}
where $m$ is the diagonal mass tensor, $\nu$ is a damping coefficient and $\Delta t$ is the time step of the discretized trajectory. Equation \eqref{eq9} has the form of a harmonic interreplica interaction with variable natural spring lengths
\begin{equation}
    \label{eq10}
    {\bf L}_i=-\frac{\Delta t}{m \nu}\nabla E({\bf R}_i).
\end{equation}
By minimizing the OM action, one generates the most probable discretized trajectory under the assumption that the underlying dynamics is Brownian. In this framework, the length of the path has a clear physical interpretation as it defines its total duration, $N\Delta t$. The problem of sampling statistically relevant dynamical trajectories based on the OM action has been addressed elsewhere (see for example Refs. \onlinecite{Mandelli2020,Fujisaki2010,Fujisaki2013,Lee2017}).

In this work, we are not interested in constructing real dynamical trajectories, which typically requires the use of hundreds of images, but rather in finding the MEP. Our modification to the NEB algorithm starts from the observation that the natural spring length ${\bf L}_i$ vanishes at the saddle points of the PES. This suggests that using equation \eqref{eq9} to define the spring forces in a NEB-like calculation would naturally lead to an accumulation of images around the saddle point. We therefore propose to substitute the spring forces, ${\bf F}_i^s$ and ${\bf F}_i^{s'}$, defined in equations \eqref{eq4} and \eqref{eq6}, with
\begin{equation}
    \label{eq11}
    {\bf F}_i^{\rm OM}=k_{\rm OM}({\bf R}_{i+1}+{\bf R}_{i-1}-2{\bf R}_{i}+{\bf L}_{i-1}-{\bf L}_{i}).
\end{equation}
Following the NEB procedure, the MEP is found by putting to zero the forces
\begin{equation}
    \label{eq12}
    {\bf F}_i=-\nabla E({\bf R}_i)|_{\perp}+{\bf F}_i^{\rm OM}|_{||}+{\bf F}_i^{\rm OM}|_{\perp}.
\end{equation}
Where the first term is defined in equation \eqref{eq3}, the second term is given by
\begin{equation}
    \label{eq13}
    {\bf F}_i^{\rm OM}|_{||}={\bf F}_i^{\rm OM}\cdot{\bf \tau}_i{\bf \tau}_i,
\end{equation}
and the last term is given by
\begin{equation}
    \label{eq14}
    {\bf F}_i^{\rm OM}|_{\perp}=f(\phi_i)({\bf F}_i^{\rm OM}-{\bf F}_i^{\rm OM}\cdot{\bf \tau}_i{\bf \tau}_i).
\end{equation}
At the end of the minimization, a final “climbing image” optimization is performed as discussed in the last paragraph of Section \ref{subsec:NEB}.

We note that, strictly speaking, equation \eqref{eq11} is not the exact gradient of the OM action since we have neglected the dependence of the natural spring length on the position. The exact gradient can be implemented using a finite difference formula at the cost of two more force computations per optimization step \cite{Mandelli2020}. Since our aim is to increase the resolution of the MEP around the saddle, equation \eqref{eq11} is already a reasonable choice, as demonstrated by our results. This avoids having to compute second derivatives of the potential energy. With this implementation, the number of force calculations per optimization step is the same as in the standard NEB algorithm. We have checked that the results do not change when using the exact gradient of equation \eqref{eq9}.

Because we are not interested in the true dynamics of the system, the spring constant $k_{\rm OM}$ loses its original dynamical interpretation and becomes a user-defined parameter. Nevertheless, we observed that better results are obtained defining the spring constant as $k_{\rm OM}=m\nu/2\Delta t$, following equation \eqref{eq9}. In all the simulations, we have used a value of $\Delta t=1$ fs, which is a typical time step adopted in atomistic molecular dynamics simulations. As a rule of thumb, we have found that a good choice of $\nu$ corresponds to values satisfying $\nu\Delta t=1$.
\subsection{\label{subsec:Models} Models and simulation protocols}
Simulations of alanine dipeptide in vacuum were performed using the amber99-SB \cite{Hornak2006} force field. Because this force field is not implemented in LAMMPS \cite{Plimpton1995}, we have used the {\it convert.py} python script that is part of the InterMol \cite{Shirts2017} software to generate input files implementing an equivalent force field. In all simulations, long-range interactions between periodic images have not been included. Path optimizations were performed fixing the C$_{\alpha}$ carbon of all the images in the origin, with the carbon atom of the methyl residue aligned along the z axis and constrained to move only along it, the hydrogen atom of the C$_{\alpha}$ carbon free to move only in the $(y,z)$ plane and all other atoms free to move in all directions. The free energy landscape of alanine dipeptide in vacuum was obtained using the On-the-fly Probability Enhanced Sampling (OPES) method \cite{Invernizzi2020}, the details of the OPES simulation can be found in Ref. \onlinecite{Invernizzi2020}.

Simulations of the dehydrogenation of ethane were performed using the C/H/O ReaxFF \cite{Chenoweth2008} force field. Path optimizations were performed fixing one carbon atom of all the images at the origin, with the second carbon atom aligned along the $x$ axis and constrained to move only along it and all other atoms free to move in all direction. The final configuration adopted in the simulations corresponds to the hydrogen molecule with its bond parallel to the carbon-carbon bond of ethylene and kept at a distance of 2.7 \AA\, away from it.

Simulations of graphene were performed using the LCBOPII \cite{Los2005} force field. The equilibrium carbon-carbon distance in graphene is C-C$_{\rm eq}=1.42$ \AA. In all simulations, we considered a rectangular supercell of sides $L_x=39.35$ \AA\, and $L_y=34.08$ \AA, containing 512 atoms, and we applied periodic boundary conditions in $x$ and $y$.  Path optimizations were performed keeping fixed a selected atom far from the defect. We checked that relaxations with all the atoms free to move in all directions lead to the same results.

All simulations were performed using LAMMPS \cite{Plimpton1995}. Structural optimizations used to generate the endpoint configurations were performed adopting the Fast Inertial Relaxation Engine (FIRE) \cite{Bitzek2006} minimization algorithm. All standard NEB calculations were performed using the LAMMPS implementation of the climbing-image NEB algorithm of Ref. \onlinecite{Henkelman2000a} adopting the FIRE \cite{Bitzek2006} minimization algorithm for structural optimizations. The modified NEB algorithm has been implemented in LAMMPS, building on our previous implementation of the OM action \cite{Mandelli2020}. Structural optimizations using our modified NEB algorithm were performed adopting a projected velocity Verlet algorithm \cite{Jonsson1998}. In all cases, optimization was stopped when the maximum force acting on the atoms of each image was at least smaller than 1 meV/\AA. The starting discretized paths were constructed using a linear interpolation between the two endpoints.
\begin{figure*}
\includegraphics[width=0.7\textwidth]{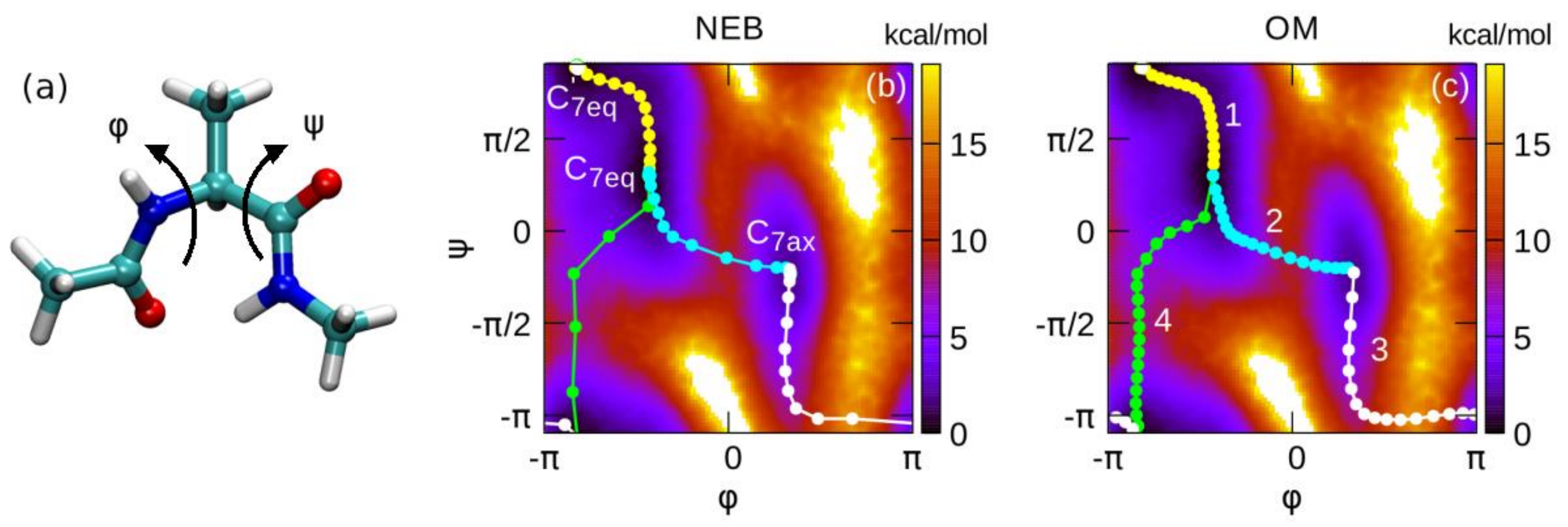}
\caption{\label{fig1} (a) Ball-and-stick model of alanine dipeptide showing the two backbone dihedral angles $\phi$ and $\psi$. Panels (b) and (c) report the four MEPs investigated obtained using (b) the standard NEB algorithm and (c) our modified version. As a reference, the free energy surface at $T=300$ K is shown as a colored map. The results reported here were obtained using $k_{||}=1$ kcal/mol/\AA$^2$, $k_{\perp}=0$ and $\nu=1$ fs$^{-1}$. In panel (b), path number 4 appears to be much coarser than the corresponding one in panel (c). The reason is that the images near the endpoints of this path differ from each other only in the angular orientation of the methyl groups of the blocked terminal ends. Since these rotations do not involve the dihedral angles $(\phi,\psi)$, these images appear to be bunched together at the two local minima.}
\end{figure*}
\subsection{\label{subsec:Definitions}Definitions}
Within the harmonic approximation to TST, the kinetic rate is given by \cite{Vineyard1957,Voter1984}
\begin{equation}
    \label{eq15}
    \tau^{-1}=\Omega e^{-\frac{\Delta E^{\dag}}{k_B T}}.
\end{equation}
Where $\Omega = \sum_{n=1}^{D}\nu_n/\sum_{n=1}^{D-1}\nu^{\rm saddle}_n$, $D$ is the number of degrees of freedom and $\nu_n$, $\nu_n^{\rm saddle}$ are the positive normal mode frequencies of the starting configuration and of the configuration corresponding to the lowest first-order saddle point connecting the potential energy basin of the initial configuration to the one of the final state. The activation energy $\Delta E^{\dagger}=E^{\rm saddle}-E^{i}$ is the difference between the potential energy at the saddle point and the potential energy at the minimum of the starting basin.

Throughout the text, we provide the values of the formation energy $\Delta E=E^f-E^i$, where $E^f$ is the energy of the final configuration, of the forward and backward activation energies $\Delta E^{\dag,\ddag}=E^{\rm saddle}-E^{i,f}$, and of the corresponding frequency prefactors $\Omega^{\dag,\ddag}$.

The profiles of the potential energy and of other quantities along the MEP are plotted as a function of the reaction coordinate. The latter is defined as the distance from the initial configuration, measured along the path, normalized with respect to the total path length.
\begin{table}[b]
\caption{\label{Table1}The values of $\phi$ and $\psi$ after geometry optimization of alanine dipeptide starting from configurations near the three local minima of the free energy surface shown in Figure \ref{fig1}(b). The last column report the potential energy of the different isomers relative to the most stable $C_{\rm7eq}$ configuration. Dihedral angles are in degrees and energies in kcal/mol.}
\begin{ruledtabular}
\begin{tabular}{cccc}
 &$\phi$&$\psi$&$\Delta E$\\
\hline
$C_{\rm 7eq}$   &-77.5 & 54.1 & 0\\
$C_{\rm 7eq}^{'}$ &-147.0 & 159.1 & 0.595\\
$C_{\rm 7ax}$   &60.2 & -40.9 & 1.421\\
\end{tabular}
\end{ruledtabular}
\end{table}
\begin{table}[b]
\caption{\label{Table2} The difference in potential energy between the initial and final states ($\Delta E$), the forward ($\Delta E^{\dag}$) and reverse ($\Delta E^{\ddag}$) activation energies, the dihedral angles $(\phi_{\rm TS},\psi_{\rm TS})$ at the transition state and the frequency prefactor ($\Omega$) of the four MEPs investigated. Paths are numbered according to Figure \ref{fig1}(c). Energies are in kcal/mol, dihedrals in degrees, frequencies in THz.}
\begin{ruledtabular}
\begin{tabular}{cccccccc}
 Path&$\Delta E$&$\Delta E^\dag$&$\Delta E^\ddag$&$\phi_{\rm TS}$&$\psi_{\rm TS}$&$\Omega^\dag$&$\Omega^\ddag$\\
\hline
1& 0.595&  1.962&  1.365&  -81.6& 120.2& 2.4& 5.9\\
2& 1.421&  8.694&  7.272&   -2.1& -26.4& 2.6& 3.6\\
3& 0.826& 14.653& 13.827&  121.2& 176.7& 4.1& 1.2\\
4& 0.595&  8.055&  7.460& -149.6& -93.6& 3.2& 1.3\\
\end{tabular}
\end{ruledtabular}
\end{table}
\section{\label{sec:Results} Results}
\subsection{\label{subsec:Alanine} Isomerization of alanine dipeptide}
As a first case study, we have considered the isomerization of alanine dipeptide in vacuum. Figure \ref{fig1}(a) shows a ball-and-stick model of the molecule along with the definition of the two backbone dihedral angles $(\phi,\psi)$ used to describe its different conformations. Stable configurations have been identified looking at the free energy landscape shown in Figure \ref{fig1}(b), which is characterized by the presence of three main basins, labelled $C_{\rm 7eq}$, $C_{\rm 7eq}^{'}$ and $C_{\rm 7ax}$. To generate the endpoint configurations for the construction of the MEPs, we have used finite temperature molecular dynamics simulations to sample configurations in each of the three metastable basins. Starting from these configurations, we have performed geometry optimizations and obtained fully relaxed structures of the three isomers. Table \ref{Table1} reports the values of the backbone dihedral angles of the optimized configurations and the corresponding potential energy measured relative to the most stable $C_{\rm 7eq}$ isomer. We have subsequently considered four paths connecting the three basins and crossing different energy barriers. For each of them, we have obtained the MEP using the standard NEB algorithm and our modified version.

Figures \ref{fig1}(b) and (c) report the results obtained using $N=20$ images per path, showing that our modified version systematically improves the resolution around the saddle points. This is particularly evident for paths number 2, 3 and 4. These paths cross higher and sharper energy barriers (see also Table \ref{Table2}). This is the typical situation where the standard NEB algorithms leads to a poor sampling of the transition region.

\begin{figure}
\includegraphics[width=0.37\textwidth]{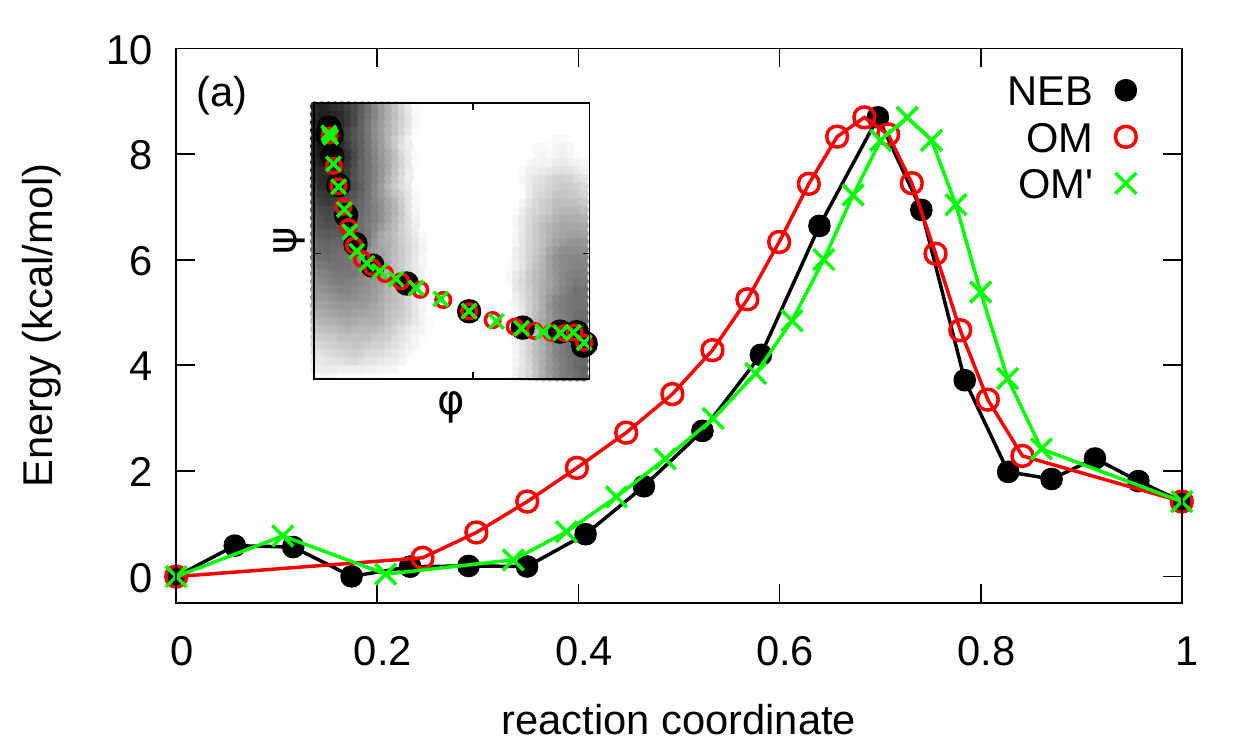}
\includegraphics[width=0.37\textwidth]{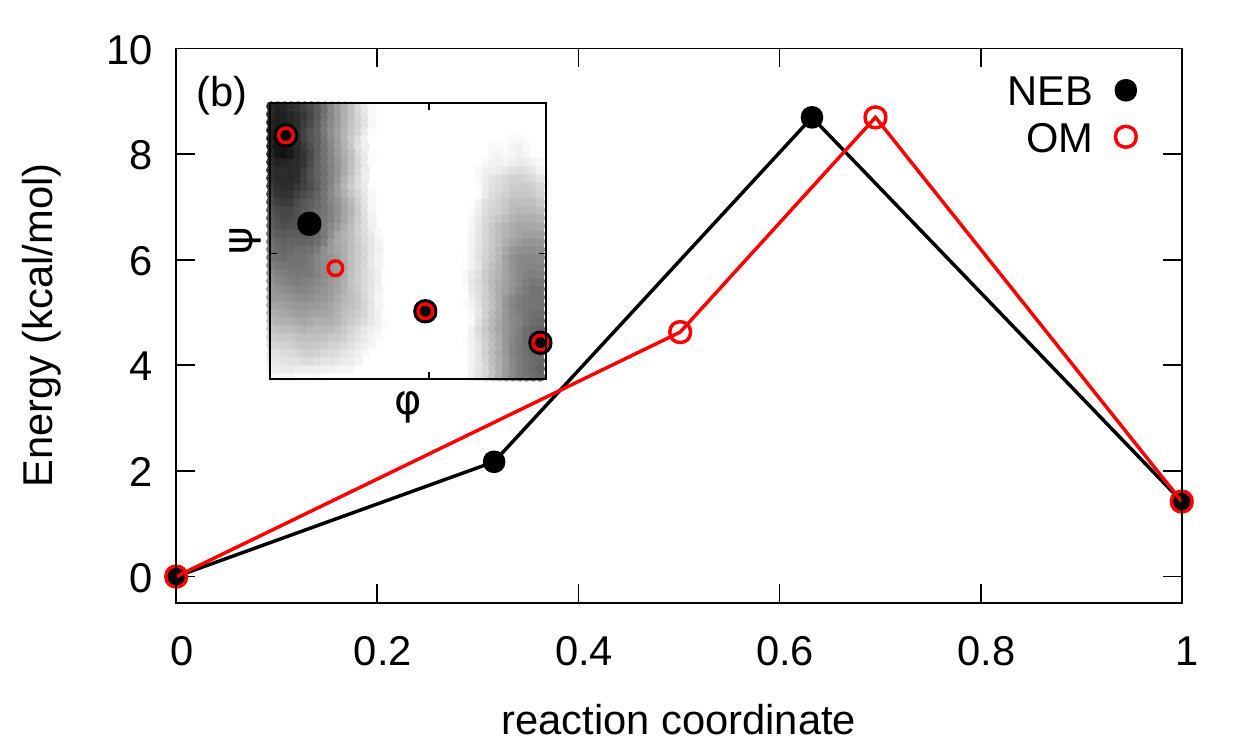}
\caption{ \label{fig2} The potential energy along path number 2, connecting the $C_{\rm 7eq}$ and the $C_{\rm 7ax}$ isomers (see Figure \ref{fig1}(b)), obtained using (a) $N=20$ and (b) $N=4$ images. The insets are zoom-in of the paths in the $(\phi,\psi)$ plane, plotted over the free energy surface. The black and the red curves correspond to the results obtained using the standard and our modified NEB algorithm, respectively. The green curves labeled OM$^{'}$ are results obtained using the exact gradient of equation \eqref{eq9} instead of the simplified expression of equation \eqref{eq11} (see discussion in Section \ref{subsec:NOM}). In panel (a), the difference between the curves at the beginning and at the end of the path is due to rotations of the methyl groups of the blocked terminal ends that are absent in the MEP obtained using the modified NEB. We note that these rotations are present at the beginning of the path when using the exact gradient of equation \eqref{eq9} (see green curve), however, this does not affect the resolution around the saddle point. The parameters used are the same as reported in the caption of Figure 1.}
\end{figure}

For all paths considered, both algorithms relax towards the same MEP and locate the same saddle point. This is made clearer in Figure \ref{fig2}(a), where we report the potential energy profile along the MEP corresponding to path number 2. In the region crossing the sharp energy barrier, roughly corresponding to values of the reaction coordinate between 0.6 and 0.8, the density of images is nearly doubled using our algorithm. In Figures S1-S3 of the Supplementary Material we report the results for paths number 1, 3 and 4.

\begin{figure}[tbh]
\includegraphics[width=0.37\textwidth]{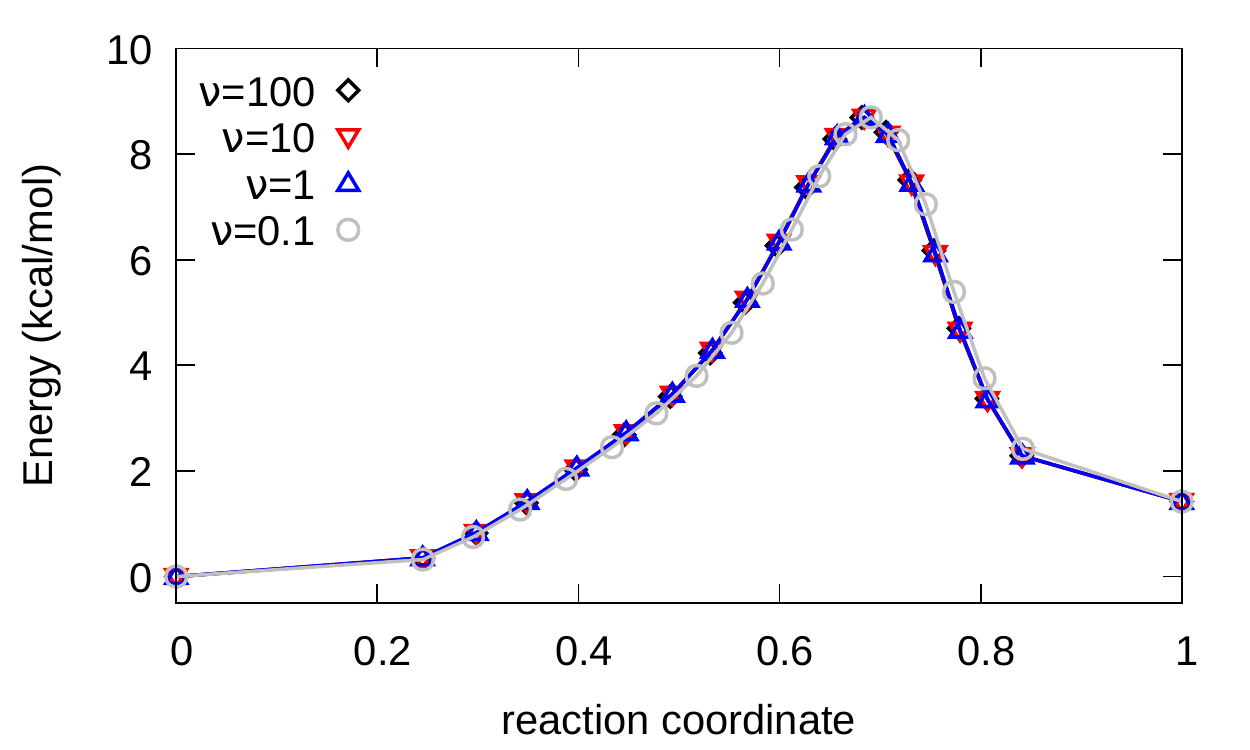}
\caption{\label{fig4} The potential energy along path number 2, obtained using the modified NEB algorithm using different values of the parameter $\nu$, as indicated in the legend (units are fs$^{-1}$).}
\end{figure}

In order to check the dependence of the results on the number of images, we performed additional simulations varying $N$ in the range $3\leq N\leq20$. In all cases, both algorithms identified the same saddle point, independent of the number of images adopted, while our modified scheme always lead to an improved resolution around the saddle (see Figures S4-S7 in the Supplementary Material). As an example, Figure \ref{fig2}(b) shows the results obtained using $N=4$ images.

Finally, in Figure \ref{fig4} we study the dependence of the results on the value of the parameter $\nu$, defining the mass-dependent spring constant $k_{\rm OM}=m\nu/2\Delta t$ in our modified scheme. Results show that the MEP is independent of the value of $\nu$ across several orders of magnitude. Similar results were obtained for all the paths considered (see Figures S8-S10 in the Supplementary Materials).

We further checked that the results are independent of the initial condition by performing standard NEB optimizations starting from the path optimized using our modified algorithm. In these tests, we never observed an increase in resolution near the saddle point and the path always relaxed to the same MEP obtained starting from the linear interpolation.
\begin{figure}[tb]
\includegraphics[width=0.37\textwidth]{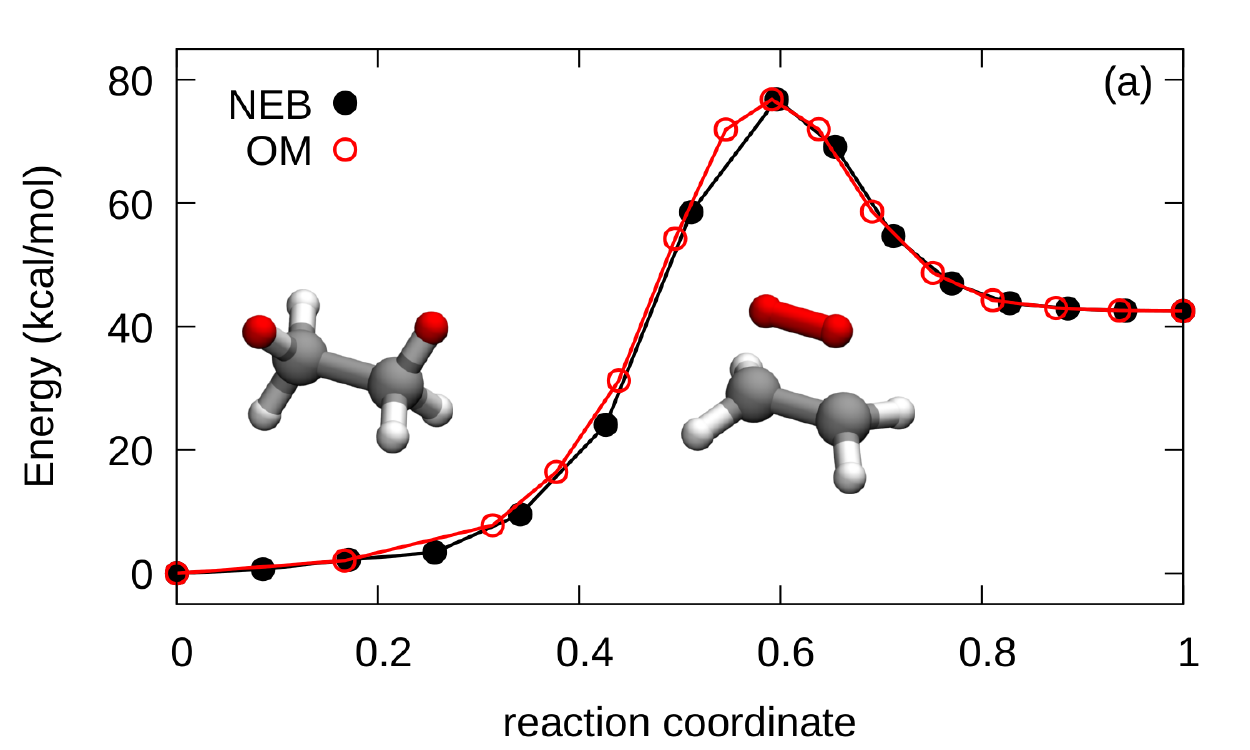}
\includegraphics[width=0.37\textwidth]{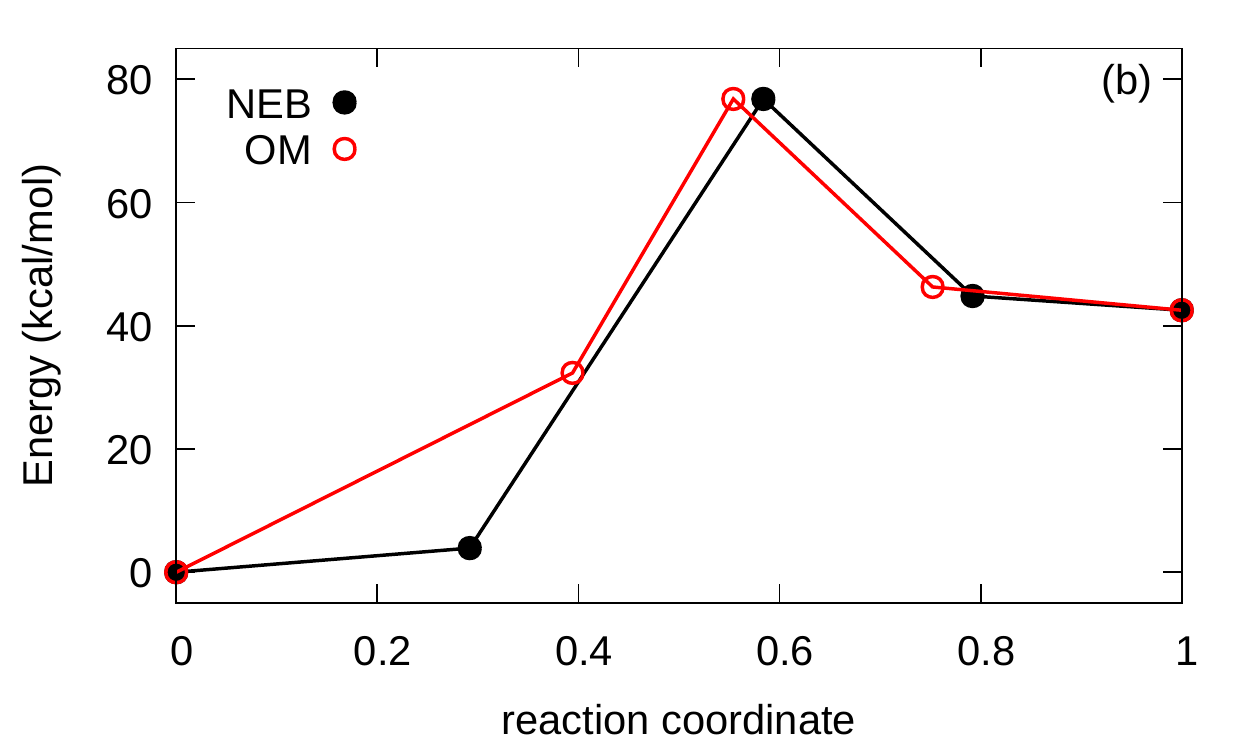}
\caption{\label{fig5} (a) The potential energy along the MEP for the dehydrogenation of ethane obtained using $N=15$ images. The black and the red curves correspond to the results obtained using the standard and the modified NEB algorithm, respectively. The insets show the ethane molecule in the initial configuration and the structure at the transition state. The hydrogen atoms that detach from ethane during the reaction are colored in red. (b) Potential energy along the MEP obtained using $N=5$ images. The results in both panels were obtained using $k_{||}=1$ kcal/mol/\AA$^2$, $k_{\perp}=0$ and $\nu=1$ fs$^{-1}$.}
\end{figure}
\begin{figure}[t]
\includegraphics[width=0.37\textwidth]{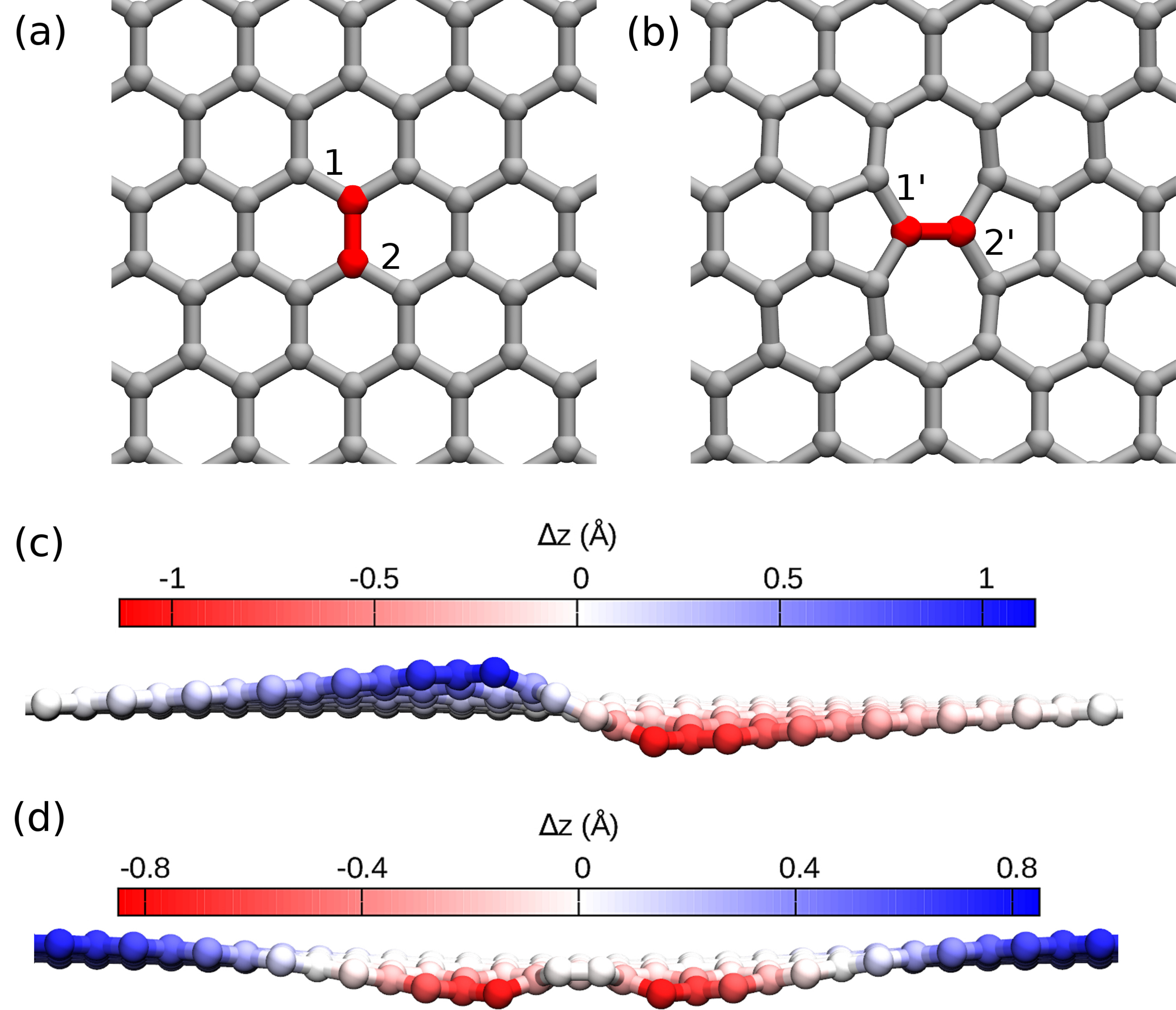}
\caption{\label{fig6} (a) Top view of perfect graphene. The bond that undergoes a ninety degrees rotation is marked in red and the two bonds that break during the transformation are also indicated. (b) Top view of an optimized, buckled SW defect. The rotated bond at the core of the defect is marked in red and the two new bonds that form during the transformation are indicated. (c) Lateral view of the sinelike buckled SW defect. (d) Lateral view of the cosinelike buckled SW defect. In panels (c) and (d), carbon atoms are colored according to their vertical position relative to the basal plane of graphene. The supercells are shown with the bond that undergoes the rotation in the foreground and in central position.}
\end{figure}
\subsection{\label{subsec:Ethane} Dehydrogenation of ethane}
As a second example, we have studied the dehydrogenation of ethane, $C_2H_6\rightleftharpoons C_2H_4+H_2$.
This reaction is an endothermic elimination reaction. With the adopted force field, the difference in energy between reactants and products is $\Delta E=43$ kcal/mol. During the reaction, two C-H bonds break and the carbon hybridization changes from $sp_3$ to $sp_2$. The dehydrogenation of light alkanes is of interest in the chemical industry for the production of chemical intermediates based on alkenes. In industrial applications, the $C-H$ bond is activated using catalyzers. Here we consider the reaction in absence of catalyzers. In this picture, the first step of the reaction consists in a rotation of the methyl groups that brings two hydrogen atoms in alignment. Subsequently, the hydrogen atoms approach, the $C-H$ bonds break, the $H_2$ molecule leaves and ethylene reaches its equilibrium planar structure. In our simulations, after the initial rotation of the methyl groups, the system maintains a symmetric configuration with the carbon-carbon and hydrogen-hydrogen bonds parallel to each other and centered. The carbon-carbon distance changes from 1.57 to 1.53 to 1.33 \AA, respectively in ethane, at the transition state and in ethylene. The forward and reverse activation barriers are $\Delta E^\dag=76.8$ kcal/mol and $\Delta E^\ddag=33.8$ kcal/mol, respectively. The frequency prefactor in equation \eqref{eq15} for the dehydrogenation is $\Omega^\dag=639$ THz.

In Figure \ref{fig5}(a) we report the MEPs obtained using the standard and modified NEB algorithm using $N=15$ images. Our method roughly doubles the density of images in the transition region of the path corresponding to the reaction coordinate in the interval $[0.5,0.7]$ around the saddle point. The optimized paths as well as the saddle points obtained from the two protocols coincide. We checked that the results of our modified algorithm are the same in the parameter range of $0.1<\nu<100$ fs$^{-1}$ (see Figure S11 in the Supplementary Materials). 

In order to check the dependence of the results on the number of images, we performed additional simulations varying $N$ in the range $5\leq N\leq15$. In all cases, both algorithms identified the same saddle point, independent of the number of images adopted. Also in this case, our modified scheme systematically leads to an improved resolution around the saddle. As an example, in Figure \ref{fig2}(b) we report the results obtained using $N=5$ images. In Figure S12 of the Supplementary Material we report the results for all the values of $N$ investigated.

\subsection{\label{subsec:Gra} Healing of a 5-77-5 defect in graphene}
An important defect in graphite is the so-called 5-77-5 topological defect, which is formed via a rotation by $\pi/2$ of a carbon-carbon bond in the graphene sheet (see Figure \ref{fig6}(a) and (b)). This rotation is also called Stone-Wales (SW) transformation \cite{Stone1986} and it involves the breaking of two pristine carbon-carbon bonds (1 and 2 in panel (a)) and the subsequent formation of two new bonds (1’ and 2’ in panel (b)). In the process, four hexagons of the original graphene honeycomb are converted into two pentagons and two heptagons. The SW defect plays an important role in the formation of fullerenes and nanotubes. In fact, the planar configuration of the SW defect is not stable and the flexible graphene sheet can easily reduce the compressive strain at the core of the defect by acquiring a buckled configuration. Two metastable geometries have been identified in the literature \cite{Ma2009}, corresponding to the sinelike and cosinelike structures. These are shown in Figure \ref{fig6}(c) and (d), as obtained after geometry optimization using our model graphene supercell. The sinelike structure is characterized by a larger buckling amplitude of $h=2.26$ \AA\, as compared to the value of $h=1.68$ \AA\, in the cosinelike geometry. On the other hand, the length of the rotating carbon-carbon bond is the same in both structures, being $\sim$4 \% compressed with respect to the equilibrium bond length of the pristine sheet (see Table \ref{Table3}). In agreement with previous results \cite{Ma2009}, we found that the sinelike configuration is energetically favored. The formation energies of the sinelike and cosinelike defects are $\Delta E=4.33$ and 4.53 eV, respectively, in fair agreement with previous density functional theory calculations of the buckled structures \cite{Ma2009} and with the values reported in Ref. \onlinecite{Los2005} for the unstable planar geometry using the same force field adopted here. In the following, we focus on the sinelike defect.
\begin{table}[t]
\caption{\label{Table3} The values of the buckling amplitude ($h$) and of the carbon-carbon bond length of the bond that undergoes the rotations in the optimized buckled sinelike and cosinelike SW defects shown in Figure \ref{fig6}. Amplitudes are in \AA\, and bond lengths are in units of the equilibrium value of the adopted force field, $C-C_{eq}=1.42$ \AA.}
\begin{ruledtabular}
\begin{tabular}{ccc}
 &$h$&$C-C$\\
\hline
Sinelike  & 2.26 & 0.964\\
Cosinelike& 1.68 & 0.965
\end{tabular}
\end{ruledtabular}
\end{table}
\begin{table}[t]
\caption{\label{Table4} The formation energy ($\Delta E$), the forward ($\Delta E^\dag$) and reverse ($\Delta E^\ddag$) activation energies, the angle ($\theta_{\rm TS}$ ), carbon-carbon bond length ($C-C_{\rm TS}$), and buckling amplitude ($h_{\rm TS}$) at the transition state and the frequency prefactor ($\Omega$) of the MEPs obtained for the sinelike and cosinelike SW defect. Energies are in eV, amplitudes in \AA, angles in degrees, bond lengths are in units of the equilibrium value ($C-C_{\rm eq}=1.42$ \AA) and frequencies are in THz.}
\begin{ruledtabular}
\begin{tabular}{ccccccccc}
 &$\Delta E$&$\Delta E^\dag$&$\Delta E^\ddag$&$h_{\rm TS}$&$\theta_{\rm TS}$&$C-C_{\rm TS}$&$\Omega^\dag$&$\Omega^\ddag$\\
\hline
Sinelike  & 4.43 & 4.64 & 8.97 & 1.68 & 46.41 & 0.871 & 6.5 & 0.05\\
Cosinelike& 4.56 & 4.40 & 8.96 & 1.17 & 39.04 & 0.872 & 5.2 & 0.07
\end{tabular}
\end{ruledtabular}
\end{table}
\begin{figure}[t]
\includegraphics[width=0.37\textwidth]{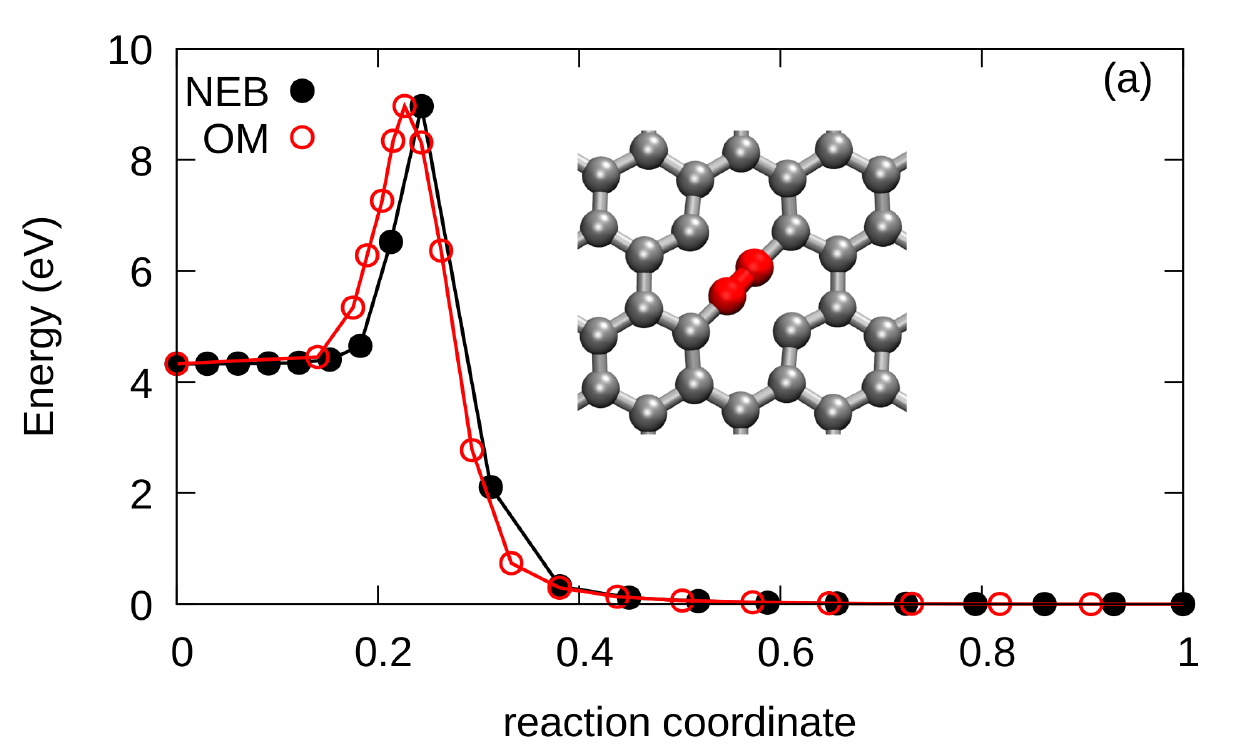}
\includegraphics[width=0.37\textwidth]{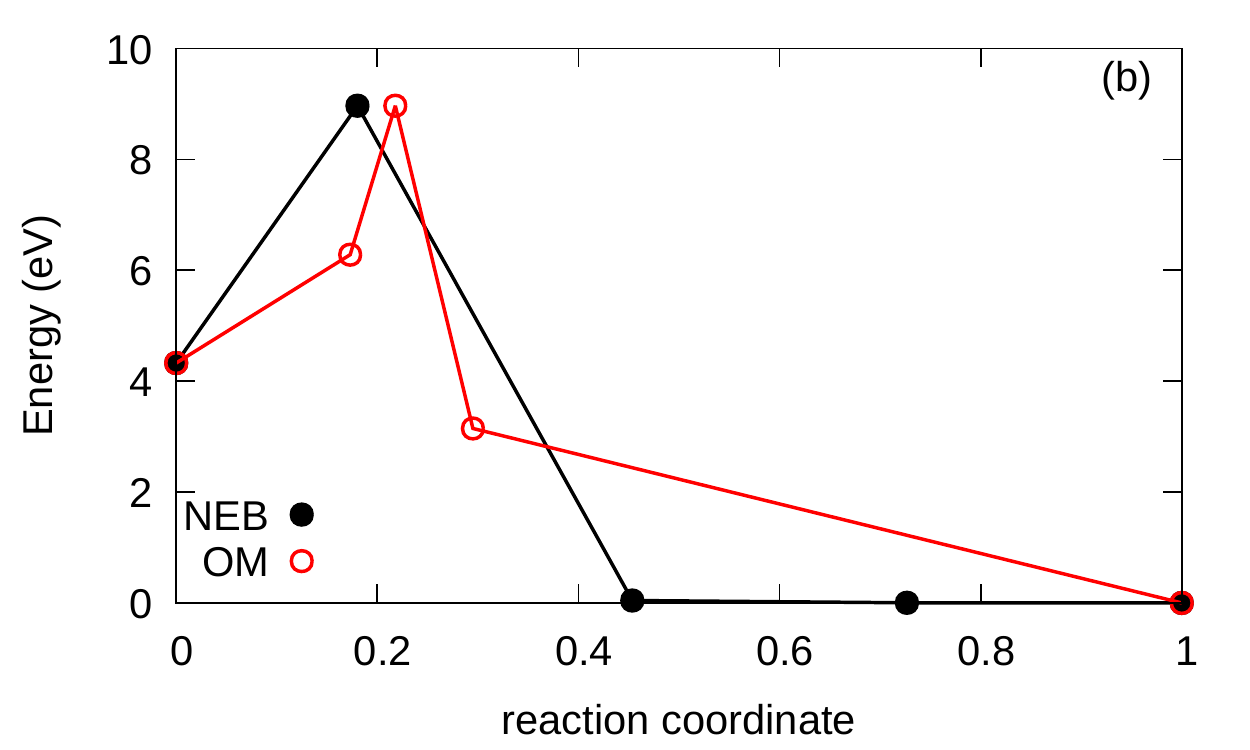}
\caption{\label{fig7} The potential energy along the MEP for the healing of the sinelike SW defect obtained using (a) $N=20$ and (b) $N=5$ images. The black and the red curves correspond to the results obtained using the standard and the modified NEB algorithm, respectively. The inset in panel (a) is a zoom-in of the region around the core of the defect at the transition state. The bond that undergoes the ninety degrees clockwise rotation is marked in red. The results reported here were obtained using $k_{||}=1$ eV/\AA$^2$, $k_{\perp}=0$ and $\nu=1$ fs$^{-1}$.}
\end{figure}
\begin{figure}[h!]
\includegraphics[width=0.5\textwidth]{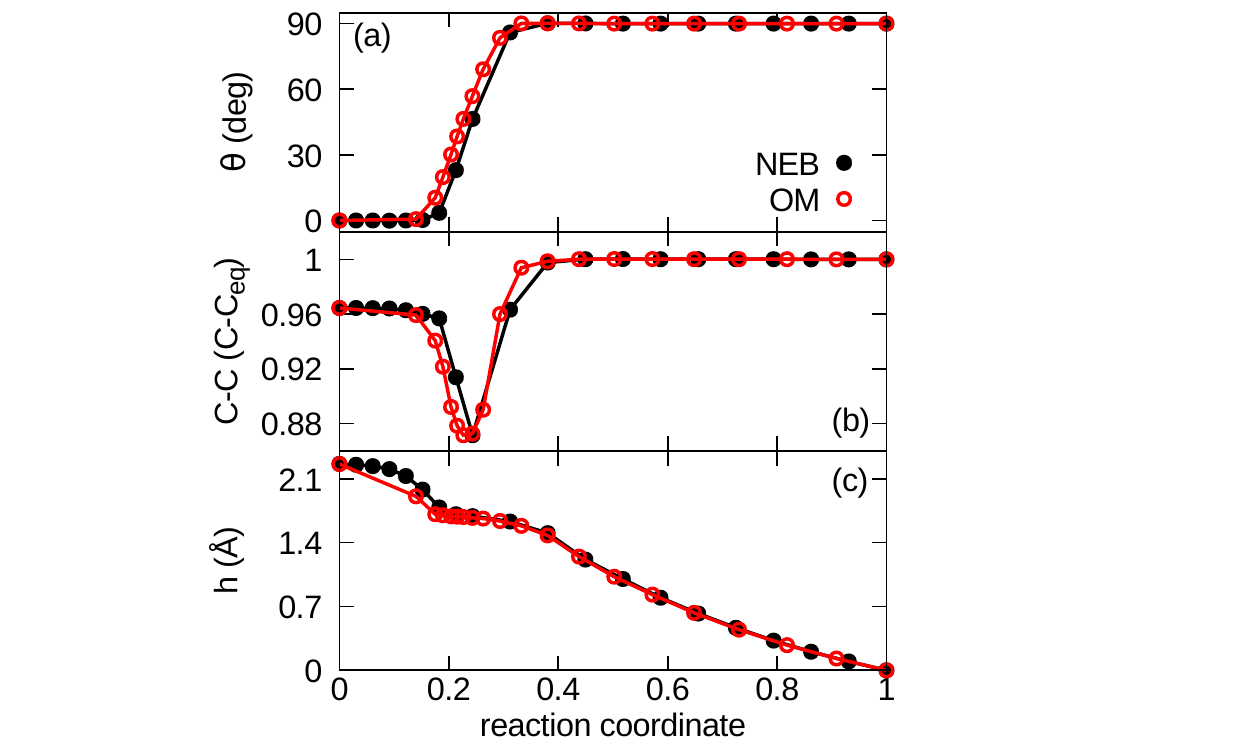}
\caption{\label{fig8} (a) The angle and (b) the carbon-carbon bond length of the bond that undergoes the ninety degrees rotation and (c) the buckling amplitude along the MEP for the healing of the sinelike SW defect. The black and the red curves correspond to the results obtained using the standard and the modified NEB algorithm, respectively, using $N=20$ images. }
\end{figure}

In Figure \ref{fig7}(a) we report the MEPs for the healing of the SW defect, obtained using $N=20$ images. Also in this case, the modified algorithm leads to an improved resolution of the MEP around the saddle point, roughly doubling the density of images in the region straddling the sharp energy peak. This has allowed us to investigate in details the structure of the core of the defect along the path. In Figure \ref{fig8} we report the profile of three geometrical parameters, namely, the angle $\theta$ and the length of the rotating bond and the overall buckling amplitude $h$. In the transition region, the angle increases linearly (see panel (a)), while the bond length shrinks, reaching a minimum value at the saddle point, and then increases towards the equilibrium value (see panel (b)). The saddle point configuration corresponds to an almost linear geometry of the four carbons at the core of the defect, suggesting an sp-like hybridization of the two rotating carbon atoms and the formation of a triple bond (see inset in Figure \ref{fig7}(a) and Table \ref{Table4}). Accordingly, the value of 1.24 \AA\, measured at the saddle point coincides with the length of the triple bond described by the adopted force field (see Table \ref{Table2} in Ref. \onlinecite{Los2005}). Along the MEP, the buckling amplitude decreases monotonously and is characterized by an almost flat plateau around the saddle (see panel (c)). For completeness, in Table \ref{Table4} we report the energetics of the reaction and the geometrical parameters characterizing the saddle points that one needs to cross to heal the sinelike and cosinelike SW defects. We checked that results obtained with our modified algorithm are the same in the range $0.1<\nu<10$ fs$^{-1}$ of the parameter $\nu$ defining the spring constant $k_{\rm OM}=m\nu/2\Delta t$ (see Figure S13 in the Supplementary Materials).

Simulations carried out using different number $N$ of images always resulted in the identification of the same saddle point, independent of $N$ and of the algorithm adopted. Also in this case, our modified scheme always lead to an improved resolution around the saddle. This is exemplified here in Figure \ref{fig7}(b), showing results for $N=5$. In Figures S14 and S15 of the Supplementary Material we report the analysis of the results obtained for all the values of $N$ investigated.

\section{\label{sec:Conclusions} Discussion and Conclusions}
In this work, we have addressed the problem of increasing the resolution around the saddle point of discretized minimum energy paths obtained within the framework of a nudged-elastic-band-like algorithm. Other approaches have already been presented in the literature. Notably, in a first modification to the original NEB algorithm \cite{Henkelman2000a}, it was suggested to automatically increase the spring constant in proximity of the saddle point. Subsequently, other variants of the NEB algorithm have been proposed, where a sequence of standard NEB calculations is performed that iteratively increase the resolution around the image with higher energy by locally adding new replicas \cite{Maragakis2002,Kolsbjerg2016}. These methods have been designed to improve computational efficiency by relaxing only few replicas around the putative saddle point at each iteration. Our method differs from these approaches as it modifies directly the expression of the forces that are minimized, while the algorithm remains the same as in the standard NEB approach \cite{Henkelman2000,Henkelman2000a} with fixed number of images. Unlike Ref. \onlinecite{Henkelman2000a}, rather than adjusting the spring constant, our method exploits the stationary property of the saddle point to define a locally adaptive natural spring length that vanishes at the saddle.

In conclusion, we have presented a modified NEB method that leads to an automatic increase in resolution around saddle points of the potential energy surface. We demonstrated the ability of the method in cases of practical interest, selected out of the realm of physical chemistry and materials science and compared it to the NEB algorithm \cite{Henkelman2000,Henkelman2000a}. In all the cases studied, our approach improved the resolution of the MEP around the saddle point. In light of the results presented here, we believe that the method represents a useful alternative tool for locating saddle points in processes with known starting and final configurations.

Finally, we note that our approach can be combined straightforwardly with the algorithms of Refs. \onlinecite{Maragakis2002,Kolsbjerg2016,Zhang2016} for improved efficiency and to have more control on the number of images as they are manually added.

\section*{Supplementary Material}
The online supplementary material includes Figures S1-S15, reporting the results of additional simulations.

\section*{Acknowledgments}
The research was supported by the European Union Grant No. ERC-2014-AdG-670227/VARMET. We also thank the NCCR MARVEL, funded by the Swiss National Science Foundation.

\section*{Data availability statement}
The data that support the findings of this study are available from the corresponding author upon reasonable request.

\bibliography{biblio_JCP_nudgedOM}

\begin{thebibliography}{36}%
\makeatletter
\providecommand \@ifxundefined [1]{%
 \@ifx{#1\undefined}
}%
\providecommand \@ifnum [1]{%
 \ifnum #1\expandafter \@firstoftwo
 \else \expandafter \@secondoftwo
 \fi
}%
\providecommand \@ifx [1]{%
 \ifx #1\expandafter \@firstoftwo
 \else \expandafter \@secondoftwo
 \fi
}%
\providecommand \natexlab [1]{#1}%
\providecommand \enquote  [1]{``#1''}%
\providecommand \bibnamefont  [1]{#1}%
\providecommand \bibfnamefont [1]{#1}%
\providecommand \citenamefont [1]{#1}%
\providecommand \href@noop [0]{\@secondoftwo}%
\providecommand \href [0]{\begingroup \@sanitize@url \@href}%
\providecommand \@href[1]{\@@startlink{#1}\@@href}%
\providecommand \@@href[1]{\endgroup#1\@@endlink}%
\providecommand \@sanitize@url [0]{\catcode `\\12\catcode `\$12\catcode
  `\&12\catcode `\#12\catcode `\^12\catcode `\_12\catcode `\%12\relax}%
\providecommand \@@startlink[1]{}%
\providecommand \@@endlink[0]{}%
\providecommand \url  [0]{\begingroup\@sanitize@url \@url }%
\providecommand \@url [1]{\endgroup\@href {#1}{\urlprefix }}%
\providecommand \urlprefix  [0]{URL }%
\providecommand \Eprint [0]{\href }%
\providecommand \doibase [0]{http://dx.doi.org/}%
\providecommand \selectlanguage [0]{\@gobble}%
\providecommand \bibinfo  [0]{\@secondoftwo}%
\providecommand \bibfield  [0]{\@secondoftwo}%
\providecommand \translation [1]{[#1]}%
\providecommand \BibitemOpen [0]{}%
\providecommand \bibitemStop [0]{}%
\providecommand \bibitemNoStop [0]{.\EOS\space}%
\providecommand \EOS [0]{\spacefactor3000\relax}%
\providecommand \BibitemShut  [1]{\csname bibitem#1\endcsname}%
\let\auto@bib@innerbib\@empty
\bibitem [{\citenamefont {Dellago}\ \emph {et~al.}(1998)\citenamefont
  {Dellago}, \citenamefont {Bolhuis}, \citenamefont {Csajka},\ and\
  \citenamefont {Chandler}}]{Dellago1998a}%
  \BibitemOpen
  \bibfield  {author} {\bibinfo {author} {\bibfnamefont {C.}~\bibnamefont
  {Dellago}}, \bibinfo {author} {\bibfnamefont {P.~G.}\ \bibnamefont
  {Bolhuis}}, \bibinfo {author} {\bibfnamefont {F.~S.}\ \bibnamefont {Csajka}},
  \ and\ \bibinfo {author} {\bibfnamefont {D.}~\bibnamefont {Chandler}},\
  }\bibfield  {title} {\enquote {\bibinfo {title} {{Transition path sampling
  and the calculation of rate constants}},}\ }\href {\doibase 10.1063/1.475562}
  {\bibfield  {journal} {\bibinfo  {journal} {The Journal of Chemical Physics}\
  }\textbf {\bibinfo {volume} {108}},\ \bibinfo {pages} {1964--1977} (\bibinfo
  {year} {1998})}\BibitemShut {NoStop}%
\bibitem [{\citenamefont {van Erp}, \citenamefont {Moroni},\ and\ \citenamefont
  {Bolhuis}(2003)}]{VanErp2003}%
  \BibitemOpen
  \bibfield  {author} {\bibinfo {author} {\bibfnamefont {T.~S.}\ \bibnamefont
  {van Erp}}, \bibinfo {author} {\bibfnamefont {D.}~\bibnamefont {Moroni}}, \
  and\ \bibinfo {author} {\bibfnamefont {P.~G.}\ \bibnamefont {Bolhuis}},\
  }\bibfield  {title} {\enquote {\bibinfo {title} {{A novel path sampling
  method for the calculation of rate constants}},}\ }\href {\doibase
  10.1063/1.1562614} {\bibfield  {journal} {\bibinfo  {journal} {The Journal of
  Chemical Physics}\ }\textbf {\bibinfo {volume} {118}},\ \bibinfo {pages}
  {7762--7774} (\bibinfo {year} {2003})}\BibitemShut {NoStop}%
\bibitem [{\citenamefont {Faradjian}\ and\ \citenamefont
  {Elber}(2004)}]{Faradjian2004}%
  \BibitemOpen
  \bibfield  {author} {\bibinfo {author} {\bibfnamefont {A.~K.}\ \bibnamefont
  {Faradjian}}\ and\ \bibinfo {author} {\bibfnamefont {R.}~\bibnamefont
  {Elber}},\ }\bibfield  {title} {\enquote {\bibinfo {title} {{Computing time
  scales from reaction coordinates by milestoning}},}\ }\href {\doibase
  10.1063/1.1738640} {\bibfield  {journal} {\bibinfo  {journal} {The Journal of
  Chemical Physics}\ }\textbf {\bibinfo {volume} {120}},\ \bibinfo {pages}
  {10880--10889} (\bibinfo {year} {2004})}\BibitemShut {NoStop}%
\bibitem [{\citenamefont {Tiwary}\ and\ \citenamefont
  {Parrinello}(2013)}]{Tiwary2013}%
  \BibitemOpen
  \bibfield  {author} {\bibinfo {author} {\bibfnamefont {P.}~\bibnamefont
  {Tiwary}}\ and\ \bibinfo {author} {\bibfnamefont {M.}~\bibnamefont
  {Parrinello}},\ }\bibfield  {title} {\enquote {\bibinfo {title} {{From
  Metadynamics to Dynamics}},}\ }\href {\doibase
  10.1103/PhysRevLett.111.230602} {\bibfield  {journal} {\bibinfo  {journal}
  {Physical Review Letters}\ }\textbf {\bibinfo {volume} {111}},\ \bibinfo
  {pages} {230602} (\bibinfo {year} {2013})}\BibitemShut {NoStop}%
\bibitem [{\citenamefont {Debnath}\ and\ \citenamefont
  {Parrinello}(2020)}]{Debnath2020}%
  \BibitemOpen
  \bibfield  {author} {\bibinfo {author} {\bibfnamefont {J.}~\bibnamefont
  {Debnath}}\ and\ \bibinfo {author} {\bibfnamefont {M.}~\bibnamefont
  {Parrinello}},\ }\bibfield  {title} {\enquote {\bibinfo {title} {{Gaussian
  Mixture-Based Enhanced Sampling for Statics and Dynamics}},}\ }\href
  {\doibase 10.1021/acs.jpclett.0c01125} {\bibfield  {journal} {\bibinfo
  {journal} {The Journal of Physical Chemistry Letters}\ }\textbf {\bibinfo
  {volume} {11}},\ \bibinfo {pages} {5076--5080} (\bibinfo {year}
  {2020})}\BibitemShut {NoStop}%
\bibitem [{\citenamefont {Mandelli}, \citenamefont {Hirshberg},\ and\
  \citenamefont {Parrinello}(2020)}]{Mandelli2020}%
  \BibitemOpen
  \bibfield  {author} {\bibinfo {author} {\bibfnamefont {D.}~\bibnamefont
  {Mandelli}}, \bibinfo {author} {\bibfnamefont {B.}~\bibnamefont {Hirshberg}},
  \ and\ \bibinfo {author} {\bibfnamefont {M.}~\bibnamefont {Parrinello}},\
  }\bibfield  {title} {\enquote {\bibinfo {title} {{Metadynamics of Paths}},}\
  }\href {\doibase 10.1103/PhysRevLett.125.026001} {\bibfield  {journal}
  {\bibinfo  {journal} {Physical Review Letters}\ }\textbf {\bibinfo {volume}
  {125}},\ \bibinfo {pages} {026001} (\bibinfo {year} {2020})}\BibitemShut
  {NoStop}%
\bibitem [{\citenamefont {Eyring}(1935)}]{Eyring1935}%
  \BibitemOpen
  \bibfield  {author} {\bibinfo {author} {\bibfnamefont {H.}~\bibnamefont
  {Eyring}},\ }\bibfield  {title} {\enquote {\bibinfo {title} {{The Activated
  Complex in Chemical Reactions}},}\ }\href {\doibase 10.1063/1.1749604}
  {\bibfield  {journal} {\bibinfo  {journal} {The Journal of Chemical Physics}\
  }\textbf {\bibinfo {volume} {3}},\ \bibinfo {pages} {107--115} (\bibinfo
  {year} {1935})}\BibitemShut {NoStop}%
\bibitem [{\citenamefont {Wigner}(1938)}]{Wigner1938}%
  \BibitemOpen
  \bibfield  {author} {\bibinfo {author} {\bibfnamefont {E.}~\bibnamefont
  {Wigner}},\ }\bibfield  {title} {\enquote {\bibinfo {title} {{The transition
  state method}},}\ }\href {\doibase 10.1039/tf9383400029} {\bibfield
  {journal} {\bibinfo  {journal} {Transactions of the Faraday Society}\
  }\textbf {\bibinfo {volume} {34}},\ \bibinfo {pages} {29} (\bibinfo {year}
  {1938})}\BibitemShut {NoStop}%
\bibitem [{\citenamefont {Keck}(2007)}]{Keck2007}%
  \BibitemOpen
  \bibfield  {author} {\bibinfo {author} {\bibfnamefont {J.~C.}\ \bibnamefont
  {Keck}},\ }\bibfield  {title} {\enquote {\bibinfo {title} {{Variational
  Theory of Reaction Rates}},}\ \ }(\bibinfo  {publisher} {WILEY‐VCH
  Verlag},\ \bibinfo {year} {2007})\ pp.\ \bibinfo {pages}
  {85--121}\BibitemShut {NoStop}%
\bibitem [{\citenamefont {Henkelman}\ and\ \citenamefont
  {J{\'{o}}nsson}(1999)}]{Henkelman1999}%
  \BibitemOpen
  \bibfield  {author} {\bibinfo {author} {\bibfnamefont {G.}~\bibnamefont
  {Henkelman}}\ and\ \bibinfo {author} {\bibfnamefont {H.}~\bibnamefont
  {J{\'{o}}nsson}},\ }\bibfield  {title} {\enquote {\bibinfo {title} {{A dimer
  method for finding saddle points on high dimensional potential surfaces using
  only first derivatives}},}\ }\href {\doibase 10.1063/1.480097} {\bibfield
  {journal} {\bibinfo  {journal} {The Journal of Chemical Physics}\ }\textbf
  {\bibinfo {volume} {111}},\ \bibinfo {pages} {7010--7022} (\bibinfo {year}
  {1999})}\BibitemShut {NoStop}%
\bibitem [{\citenamefont {Jay}\ \emph {et~al.}(2020)\citenamefont {Jay},
  \citenamefont {Huet}, \citenamefont {Salles}, \citenamefont {Gunde},
  \citenamefont {Martin-Samos}, \citenamefont {Richard}, \citenamefont {Landa},
  \citenamefont {Goiffon}, \citenamefont {{De Gironcoli}}, \citenamefont
  {H{\'{e}}meryck},\ and\ \citenamefont {Mousseau}}]{Jay2020}%
  \BibitemOpen
  \bibfield  {author} {\bibinfo {author} {\bibfnamefont {A.}~\bibnamefont
  {Jay}}, \bibinfo {author} {\bibfnamefont {C.}~\bibnamefont {Huet}}, \bibinfo
  {author} {\bibfnamefont {N.}~\bibnamefont {Salles}}, \bibinfo {author}
  {\bibfnamefont {M.}~\bibnamefont {Gunde}}, \bibinfo {author} {\bibfnamefont
  {L.}~\bibnamefont {Martin-Samos}}, \bibinfo {author} {\bibfnamefont
  {N.}~\bibnamefont {Richard}}, \bibinfo {author} {\bibfnamefont
  {G.}~\bibnamefont {Landa}}, \bibinfo {author} {\bibfnamefont
  {V.}~\bibnamefont {Goiffon}}, \bibinfo {author} {\bibfnamefont
  {S.}~\bibnamefont {{De Gironcoli}}}, \bibinfo {author} {\bibfnamefont
  {A.}~\bibnamefont {H{\'{e}}meryck}}, \ and\ \bibinfo {author} {\bibfnamefont
  {N.}~\bibnamefont {Mousseau}},\ }\bibfield  {title} {\enquote {\bibinfo
  {title} {{Finding Reaction Pathways and Transition States: r-ARTn and d-ARTn
  as an Efficient and Versatile Alternative to String Approaches}},}\ }\href
  {\doibase 10.1021/acs.jctc.0c00541} {\bibfield  {journal} {\bibinfo
  {journal} {Journal of Chemical Theory and Computation}\ }\textbf {\bibinfo
  {volume} {16}},\ \bibinfo {pages} {6726--6734} (\bibinfo {year}
  {2020})}\BibitemShut {NoStop}%
\bibitem [{\citenamefont {Jonsson}, \citenamefont {Mills},\ and\ \citenamefont
  {Jacobsen}(1998)}]{Jonsson1998}%
  \BibitemOpen
  \bibfield  {author} {\bibinfo {author} {\bibfnamefont {H.}~\bibnamefont
  {Jonsson}}, \bibinfo {author} {\bibfnamefont {G.}~\bibnamefont {Mills}}, \
  and\ \bibinfo {author} {\bibfnamefont {K.~W.}\ \bibnamefont {Jacobsen}},\
  }\bibfield  {title} {\enquote {\bibinfo {title} {{Nudged elastic band method
  for finding minimum energy paths of transitions}},}\ }in\ \href {\doibase
  10.1142/9789812839664_0016} {\emph {\bibinfo {booktitle} {Classical and
  Quantum Dynamics in Condensed Phase Simulations}}}\ (\bibinfo  {publisher}
  {World Scientific},\ \bibinfo {year} {1998})\ pp.\ \bibinfo {pages}
  {385--404}\BibitemShut {NoStop}%
\bibitem [{\citenamefont {Peters}\ \emph {et~al.}(2004)\citenamefont {Peters},
  \citenamefont {Heyden}, \citenamefont {Bell},\ and\ \citenamefont
  {Chakraborty}}]{Peters2004}%
  \BibitemOpen
  \bibfield  {author} {\bibinfo {author} {\bibfnamefont {B.}~\bibnamefont
  {Peters}}, \bibinfo {author} {\bibfnamefont {A.}~\bibnamefont {Heyden}},
  \bibinfo {author} {\bibfnamefont {A.~T.}\ \bibnamefont {Bell}}, \ and\
  \bibinfo {author} {\bibfnamefont {A.}~\bibnamefont {Chakraborty}},\
  }\bibfield  {title} {\enquote {\bibinfo {title} {{A growing string method for
  determining transition states: Comparison to the nudged elastic band and
  string methods}},}\ }\href {\doibase 10.1063/1.1691018} {\bibfield  {journal}
  {\bibinfo  {journal} {Journal of Chemical Physics}\ }\textbf {\bibinfo
  {volume} {120}},\ \bibinfo {pages} {7877--7886} (\bibinfo {year}
  {2004})}\BibitemShut {NoStop}%
\bibitem [{\citenamefont {E}\ and\ \citenamefont
  {Vanden-Eijnden}(2010)}]{E2010}%
  \BibitemOpen
  \bibfield  {author} {\bibinfo {author} {\bibfnamefont {W.}~\bibnamefont {E}}\
  and\ \bibinfo {author} {\bibfnamefont {E.}~\bibnamefont {Vanden-Eijnden}},\
  }\bibfield  {title} {\enquote {\bibinfo {title} {{Transition-Path Theory and
  Path-Finding Algorithms for the Study of Rare Events}},}\ }\href {\doibase
  10.1146/annurev.physchem.040808.090412} {\bibfield  {journal} {\bibinfo
  {journal} {Annual Review of Physical Chemistry}\ }\textbf {\bibinfo {volume}
  {61}},\ \bibinfo {pages} {391--420} (\bibinfo {year} {2010})}\BibitemShut
  {NoStop}%
\bibitem [{\citenamefont {Henkelman}\ and\ \citenamefont
  {J{\'{o}}nsson}(2000)}]{Henkelman2000}%
  \BibitemOpen
  \bibfield  {author} {\bibinfo {author} {\bibfnamefont {G.}~\bibnamefont
  {Henkelman}}\ and\ \bibinfo {author} {\bibfnamefont {H.}~\bibnamefont
  {J{\'{o}}nsson}},\ }\bibfield  {title} {\enquote {\bibinfo {title} {{Improved
  tangent estimate in the nudged elastic band method for finding minimum energy
  paths and saddle points}},}\ }\href {\doibase 10.1063/1.1323224} {\bibfield
  {journal} {\bibinfo  {journal} {The Journal of Chemical Physics}\ }\textbf
  {\bibinfo {volume} {113}},\ \bibinfo {pages} {9978--9985} (\bibinfo {year}
  {2000})}\BibitemShut {NoStop}%
\bibitem [{\citenamefont {Henkelman}, \citenamefont {Uberuaga},\ and\
  \citenamefont {J{\'{o}}nsson}(2000)}]{Henkelman2000a}%
  \BibitemOpen
  \bibfield  {author} {\bibinfo {author} {\bibfnamefont {G.}~\bibnamefont
  {Henkelman}}, \bibinfo {author} {\bibfnamefont {B.~P.}\ \bibnamefont
  {Uberuaga}}, \ and\ \bibinfo {author} {\bibfnamefont {H.}~\bibnamefont
  {J{\'{o}}nsson}},\ }\bibfield  {title} {\enquote {\bibinfo {title} {{A
  climbing image nudged elastic band method for finding saddle points and
  minimum energy paths}},}\ }\href {\doibase 10.1063/1.1329672} {\bibfield
  {journal} {\bibinfo  {journal} {The Journal of Chemical Physics}\ }\textbf
  {\bibinfo {volume} {113}},\ \bibinfo {pages} {9901--9904} (\bibinfo {year}
  {2000})}\BibitemShut {NoStop}%
\bibitem [{\citenamefont {Maragakis}\ \emph {et~al.}(2002)\citenamefont
  {Maragakis}, \citenamefont {Andreev}, \citenamefont {Brumer}, \citenamefont
  {Reichman},\ and\ \citenamefont {Kaxiras}}]{Maragakis2002}%
  \BibitemOpen
  \bibfield  {author} {\bibinfo {author} {\bibfnamefont {P.}~\bibnamefont
  {Maragakis}}, \bibinfo {author} {\bibfnamefont {S.~A.}\ \bibnamefont
  {Andreev}}, \bibinfo {author} {\bibfnamefont {Y.}~\bibnamefont {Brumer}},
  \bibinfo {author} {\bibfnamefont {D.~R.}\ \bibnamefont {Reichman}}, \ and\
  \bibinfo {author} {\bibfnamefont {E.}~\bibnamefont {Kaxiras}},\ }\bibfield
  {title} {\enquote {\bibinfo {title} {{Adaptive nudged elastic band approach
  for transition state calculation}},}\ }\href {\doibase 10.1063/1.1495401}
  {\bibfield  {journal} {\bibinfo  {journal} {Journal of Chemical Physics}\
  }\textbf {\bibinfo {volume} {117}},\ \bibinfo {pages} {4651--4658} (\bibinfo
  {year} {2002})}\BibitemShut {NoStop}%
\bibitem [{\citenamefont {Kolsbjerg}, \citenamefont {Groves},\ and\
  \citenamefont {Hammer}(2016)}]{Kolsbjerg2016}%
  \BibitemOpen
  \bibfield  {author} {\bibinfo {author} {\bibfnamefont {E.~L.}\ \bibnamefont
  {Kolsbjerg}}, \bibinfo {author} {\bibfnamefont {M.~N.}\ \bibnamefont
  {Groves}}, \ and\ \bibinfo {author} {\bibfnamefont {B.}~\bibnamefont
  {Hammer}},\ }\bibfield  {title} {\enquote {\bibinfo {title} {{An automated
  nudged elastic band method}},}\ }\href {\doibase 10.1063/1.4961868}
  {\bibfield  {journal} {\bibinfo  {journal} {Journal of Chemical Physics}\
  }\textbf {\bibinfo {volume} {145}},\ \bibinfo {pages} {094107} (\bibinfo
  {year} {2016})}\BibitemShut {NoStop}%
\bibitem [{\citenamefont {Onsager}\ and\ \citenamefont
  {Machlup}(1953)}]{Onsager1953}%
  \BibitemOpen
  \bibfield  {author} {\bibinfo {author} {\bibfnamefont {L.}~\bibnamefont
  {Onsager}}\ and\ \bibinfo {author} {\bibfnamefont {S.}~\bibnamefont
  {Machlup}},\ }\bibfield  {title} {\enquote {\bibinfo {title} {{Fluctuations
  and Irreversible Processes}},}\ }\href {\doibase 10.1103/PhysRev.91.1505}
  {\bibfield  {journal} {\bibinfo  {journal} {Physical Review}\ }\textbf
  {\bibinfo {volume} {91}},\ \bibinfo {pages} {1505--1512} (\bibinfo {year}
  {1953})}\BibitemShut {NoStop}%
\bibitem [{\citenamefont {Maras}\ \emph {et~al.}(2016)\citenamefont {Maras},
  \citenamefont {Trushin}, \citenamefont {Stukowski}, \citenamefont
  {Ala-Nissila},\ and\ \citenamefont {J\'onsson}}]{Maras2016}%
  \BibitemOpen
  \bibfield  {author} {\bibinfo {author} {\bibfnamefont {E.}~\bibnamefont
  {Maras}}, \bibinfo {author} {\bibfnamefont {O.}~\bibnamefont {Trushin}},
  \bibinfo {author} {\bibfnamefont {A.}~\bibnamefont {Stukowski}}, \bibinfo
  {author} {\bibfnamefont {T.}~\bibnamefont {Ala-Nissila}}, \ and\ \bibinfo
  {author} {\bibfnamefont {H.}~\bibnamefont {J\'onsson}},\ }\bibfield  {title}
  {\enquote {\bibinfo {title} {{Global transition path search for dislocation
  formation in Ge on Si(001)}},}\ }\href {\doibase 10.1016/j.cpc.2016.04.001}
  {\bibfield  {journal} {\bibinfo  {journal} {Computer Physiscs
  Communications}\ }\textbf {\bibinfo {volume} {205}},\ \bibinfo {pages}
  {13--21} (\bibinfo {year} {2016})}\BibitemShut {NoStop}%
\bibitem [{\citenamefont {{\'{A}}sgeirsson}\ \emph {et~al.}(2021)\citenamefont
  {{\'{A}}sgeirsson}, \citenamefont {Birgisson}, \citenamefont {Bjornsson},
  \citenamefont {Becker}, \citenamefont {Neese}, \citenamefont {Riplinger},\
  and\ \citenamefont {J{\'{o}}nsson}}]{Asgeirsson2021}%
  \BibitemOpen
  \bibfield  {author} {\bibinfo {author} {\bibfnamefont {V.}~\bibnamefont
  {{\'{A}}sgeirsson}}, \bibinfo {author} {\bibfnamefont {B.~O.}\ \bibnamefont
  {Birgisson}}, \bibinfo {author} {\bibfnamefont {R.}~\bibnamefont
  {Bjornsson}}, \bibinfo {author} {\bibfnamefont {U.}~\bibnamefont {Becker}},
  \bibinfo {author} {\bibfnamefont {F.}~\bibnamefont {Neese}}, \bibinfo
  {author} {\bibfnamefont {C.}~\bibnamefont {Riplinger}}, \ and\ \bibinfo
  {author} {\bibfnamefont {H.}~\bibnamefont {J{\'{o}}nsson}},\ }\bibfield
  {title} {\enquote {\bibinfo {title} {{Nudged Elastic Band Method for
  Molecular Reactions Using Energy-Weighted Springs Combined with Eigenvector
  Following}},}\ }\href {\doibase 10.1021/acs.jctc.1c00462} {\bibfield
  {journal} {\bibinfo  {journal} {Journal of Chemical Theory and Computation}\
  }\textbf {\bibinfo {volume} {17}},\ \bibinfo {pages} {4929--4945} (\bibinfo
  {year} {2021})}\BibitemShut {NoStop}%
\bibitem [{\citenamefont {Fujisaki}, \citenamefont {Shiga},\ and\ \citenamefont
  {Kidera}(2010)}]{Fujisaki2010}%
  \BibitemOpen
  \bibfield  {author} {\bibinfo {author} {\bibfnamefont {H.}~\bibnamefont
  {Fujisaki}}, \bibinfo {author} {\bibfnamefont {M.}~\bibnamefont {Shiga}}, \
  and\ \bibinfo {author} {\bibfnamefont {A.}~\bibnamefont {Kidera}},\
  }\bibfield  {title} {\enquote {\bibinfo {title} {{Onsager–Machlup
  action-based path sampling and its combination with replica exchange for
  diffusive and multiple pathways}},}\ }\href {\doibase 10.1063/1.3372802}
  {\bibfield  {journal} {\bibinfo  {journal} {The Journal of Chemical Physics}\
  }\textbf {\bibinfo {volume} {132}},\ \bibinfo {pages} {134101} (\bibinfo
  {year} {2010})}\BibitemShut {NoStop}%
\bibitem [{\citenamefont {Fujisaki}\ \emph {et~al.}(2013)\citenamefont
  {Fujisaki}, \citenamefont {Shiga}, \citenamefont {Moritsugu},\ and\
  \citenamefont {Kidera}}]{Fujisaki2013}%
  \BibitemOpen
  \bibfield  {author} {\bibinfo {author} {\bibfnamefont {H.}~\bibnamefont
  {Fujisaki}}, \bibinfo {author} {\bibfnamefont {M.}~\bibnamefont {Shiga}},
  \bibinfo {author} {\bibfnamefont {K.}~\bibnamefont {Moritsugu}}, \ and\
  \bibinfo {author} {\bibfnamefont {A.}~\bibnamefont {Kidera}},\ }\bibfield
  {title} {\enquote {\bibinfo {title} {{Multiscale enhanced path sampling based
  on the Onsager-Machlup action: Application to a model polymer}},}\ }\href
  {\doibase 10.1063/1.4817209} {\bibfield  {journal} {\bibinfo  {journal} {The
  Journal of Chemical Physics}\ }\textbf {\bibinfo {volume} {139}},\ \bibinfo
  {pages} {054117} (\bibinfo {year} {2013})}\BibitemShut {NoStop}%
\bibitem [{\citenamefont {Lee}\ \emph {et~al.}(2017)\citenamefont {Lee},
  \citenamefont {Lee}, \citenamefont {Joung}, \citenamefont {Lee},\ and\
  \citenamefont {Brooks}}]{Lee2017}%
  \BibitemOpen
  \bibfield  {author} {\bibinfo {author} {\bibfnamefont {J.}~\bibnamefont
  {Lee}}, \bibinfo {author} {\bibfnamefont {I.-H.}\ \bibnamefont {Lee}},
  \bibinfo {author} {\bibfnamefont {I.}~\bibnamefont {Joung}}, \bibinfo
  {author} {\bibfnamefont {J.}~\bibnamefont {Lee}}, \ and\ \bibinfo {author}
  {\bibfnamefont {B.~R.}\ \bibnamefont {Brooks}},\ }\bibfield  {title}
  {\enquote {\bibinfo {title} {{Finding multiple reaction pathways via global
  optimization of action}},}\ }\href {\doibase 10.1038/ncomms15443} {\bibfield
  {journal} {\bibinfo  {journal} {Nature Communications}\ }\textbf {\bibinfo
  {volume} {8}},\ \bibinfo {pages} {15443} (\bibinfo {year}
  {2017})}\BibitemShut {NoStop}%
\bibitem [{\citenamefont {Hornak}\ \emph {et~al.}(2006)\citenamefont {Hornak},
  \citenamefont {Abel}, \citenamefont {Okur}, \citenamefont {Strockbine},
  \citenamefont {Roitberg},\ and\ \citenamefont {Simmerling}}]{Hornak2006}%
  \BibitemOpen
  \bibfield  {author} {\bibinfo {author} {\bibfnamefont {V.}~\bibnamefont
  {Hornak}}, \bibinfo {author} {\bibfnamefont {R.}~\bibnamefont {Abel}},
  \bibinfo {author} {\bibfnamefont {A.}~\bibnamefont {Okur}}, \bibinfo {author}
  {\bibfnamefont {B.}~\bibnamefont {Strockbine}}, \bibinfo {author}
  {\bibfnamefont {A.}~\bibnamefont {Roitberg}}, \ and\ \bibinfo {author}
  {\bibfnamefont {C.}~\bibnamefont {Simmerling}},\ }\bibfield  {title}
  {\enquote {\bibinfo {title} {{Comparison of multiple Amber force fields and
  development of improved protein backbone parameters}},}\ }\href {\doibase
  10.1002/prot.21123} {\bibfield  {journal} {\bibinfo  {journal} {Proteins:
  Structure, Function, and Bioinformatics}\ }\textbf {\bibinfo {volume} {65}},\
  \bibinfo {pages} {712--725} (\bibinfo {year} {2006})}\BibitemShut {NoStop}%
\bibitem [{\citenamefont {Plimpton}(1995)}]{Plimpton1995}%
  \BibitemOpen
  \bibfield  {author} {\bibinfo {author} {\bibfnamefont {S.}~\bibnamefont
  {Plimpton}},\ }\bibfield  {title} {\enquote {\bibinfo {title} {{Fast Parallel
  Algorithms for Short-Range Molecular Dynamics}},}\ }\href {\doibase
  10.1006/jcph.1995.1039} {\bibfield  {journal} {\bibinfo  {journal} {Journal
  of Computational Physics}\ }\textbf {\bibinfo {volume} {117}},\ \bibinfo
  {pages} {1--19} (\bibinfo {year} {1995})}\BibitemShut {NoStop}%
\bibitem [{\citenamefont {Shirts}\ \emph {et~al.}(2017)\citenamefont {Shirts},
  \citenamefont {Klein}, \citenamefont {Swails}, \citenamefont {Yin},
  \citenamefont {Gilson}, \citenamefont {Mobley}, \citenamefont {Case},\ and\
  \citenamefont {Zhong}}]{Shirts2017}%
  \BibitemOpen
  \bibfield  {author} {\bibinfo {author} {\bibfnamefont {M.~R.}\ \bibnamefont
  {Shirts}}, \bibinfo {author} {\bibfnamefont {C.}~\bibnamefont {Klein}},
  \bibinfo {author} {\bibfnamefont {J.~M.}\ \bibnamefont {Swails}}, \bibinfo
  {author} {\bibfnamefont {J.}~\bibnamefont {Yin}}, \bibinfo {author}
  {\bibfnamefont {M.~K.}\ \bibnamefont {Gilson}}, \bibinfo {author}
  {\bibfnamefont {D.~L.}\ \bibnamefont {Mobley}}, \bibinfo {author}
  {\bibfnamefont {D.~A.}\ \bibnamefont {Case}}, \ and\ \bibinfo {author}
  {\bibfnamefont {E.~D.}\ \bibnamefont {Zhong}},\ }\bibfield  {title} {\enquote
  {\bibinfo {title} {{Lessons learned from comparing molecular dynamics engines
  on the SAMPL5 dataset}},}\ }\href {\doibase 10.1007/s10822-016-9977-1}
  {\bibfield  {journal} {\bibinfo  {journal} {Journal of Computer-Aided
  Molecular Design}\ }\textbf {\bibinfo {volume} {31}},\ \bibinfo {pages}
  {147--161} (\bibinfo {year} {2017})}\BibitemShut {NoStop}%
\bibitem [{\citenamefont {Invernizzi}\ and\ \citenamefont
  {Parrinello}(2020)}]{Invernizzi2020}%
  \BibitemOpen
  \bibfield  {author} {\bibinfo {author} {\bibfnamefont {M.}~\bibnamefont
  {Invernizzi}}\ and\ \bibinfo {author} {\bibfnamefont {M.}~\bibnamefont
  {Parrinello}},\ }\bibfield  {title} {\enquote {\bibinfo {title} {{Rethinking
  Metadynamics: From Bias Potentials to Probability Distributions}},}\ }\href
  {\doibase 10.1021/acs.jpclett.0c00497} {\bibfield  {journal} {\bibinfo
  {journal} {The Journal of Physical Chemistry Letters}\ }\textbf {\bibinfo
  {volume} {11}},\ \bibinfo {pages} {2731--2736} (\bibinfo {year}
  {2020})}\BibitemShut {NoStop}%
\bibitem [{\citenamefont {Chenoweth}, \citenamefont {van Duin},\ and\
  \citenamefont {Goddard}(2008)}]{Chenoweth2008}%
  \BibitemOpen
  \bibfield  {author} {\bibinfo {author} {\bibfnamefont {K.}~\bibnamefont
  {Chenoweth}}, \bibinfo {author} {\bibfnamefont {A.~C.~T.}\ \bibnamefont {van
  Duin}}, \ and\ \bibinfo {author} {\bibfnamefont {W.~A.}\ \bibnamefont
  {Goddard}},\ }\bibfield  {title} {\enquote {\bibinfo {title} {{ReaxFF
  Reactive Force Field for Molecular Dynamics Simulations of Hydrocarbon
  Oxidation}},}\ }\href {\doibase 10.1021/jp709896w} {\bibfield  {journal}
  {\bibinfo  {journal} {The Journal of Physical Chemistry A}\ }\textbf
  {\bibinfo {volume} {112}},\ \bibinfo {pages} {1040--1053} (\bibinfo {year}
  {2008})}\BibitemShut {NoStop}%
\bibitem [{\citenamefont {Los}\ \emph {et~al.}(2005)\citenamefont {Los},
  \citenamefont {Ghiringhelli}, \citenamefont {Meijer},\ and\ \citenamefont
  {Fasolino}}]{Los2005}%
  \BibitemOpen
  \bibfield  {author} {\bibinfo {author} {\bibfnamefont {J.~H.}\ \bibnamefont
  {Los}}, \bibinfo {author} {\bibfnamefont {L.~M.}\ \bibnamefont
  {Ghiringhelli}}, \bibinfo {author} {\bibfnamefont {E.~J.}\ \bibnamefont
  {Meijer}}, \ and\ \bibinfo {author} {\bibfnamefont {A.}~\bibnamefont
  {Fasolino}},\ }\bibfield  {title} {\enquote {\bibinfo {title} {{Improved
  long-range reactive bond-order potential for carbon. I. Construction}},}\
  }\href {\doibase 10.1103/PhysRevB.72.214102} {\bibfield  {journal} {\bibinfo
  {journal} {Physical Review B}\ }\textbf {\bibinfo {volume} {72}},\ \bibinfo
  {pages} {214102} (\bibinfo {year} {2005})}\BibitemShut {NoStop}%
\bibitem [{\citenamefont {Bitzek}\ \emph {et~al.}(2006)\citenamefont {Bitzek},
  \citenamefont {Koskinen}, \citenamefont {G{\"{a}}hler}, \citenamefont
  {Moseler},\ and\ \citenamefont {Gumbsch}}]{Bitzek2006}%
  \BibitemOpen
  \bibfield  {author} {\bibinfo {author} {\bibfnamefont {E.}~\bibnamefont
  {Bitzek}}, \bibinfo {author} {\bibfnamefont {P.}~\bibnamefont {Koskinen}},
  \bibinfo {author} {\bibfnamefont {F.}~\bibnamefont {G{\"{a}}hler}}, \bibinfo
  {author} {\bibfnamefont {M.}~\bibnamefont {Moseler}}, \ and\ \bibinfo
  {author} {\bibfnamefont {P.}~\bibnamefont {Gumbsch}},\ }\bibfield  {title}
  {\enquote {\bibinfo {title} {{Structural Relaxation Made Simple}},}\ }\href
  {\doibase 10.1103/PhysRevLett.97.170201} {\bibfield  {journal} {\bibinfo
  {journal} {Physical Review Letters}\ }\textbf {\bibinfo {volume} {97}},\
  \bibinfo {pages} {170201} (\bibinfo {year} {2006})}\BibitemShut {NoStop}%
\bibitem [{\citenamefont {Vineyard}(1957)}]{Vineyard1957}%
  \BibitemOpen
  \bibfield  {author} {\bibinfo {author} {\bibfnamefont {G.~H.}\ \bibnamefont
  {Vineyard}},\ }\bibfield  {title} {\enquote {\bibinfo {title} {{Frequency
  factors and isotope effects in solid state rate processes}},}\ }\href
  {\doibase 10.1016/0022-3697(57)90059-8} {\bibfield  {journal} {\bibinfo
  {journal} {Journal of Physics and Chemistry of Solids}\ }\textbf {\bibinfo
  {volume} {3}},\ \bibinfo {pages} {121--127} (\bibinfo {year}
  {1957})}\BibitemShut {NoStop}%
\bibitem [{\citenamefont {Voter}\ and\ \citenamefont {Doll}(1984)}]{Voter1984}%
  \BibitemOpen
  \bibfield  {author} {\bibinfo {author} {\bibfnamefont {A.~F.}\ \bibnamefont
  {Voter}}\ and\ \bibinfo {author} {\bibfnamefont {J.~D.}\ \bibnamefont
  {Doll}},\ }\bibfield  {title} {\enquote {\bibinfo {title} {{Transition state
  theory description of surface self-diffusion: Comparison with classical
  trajectory results}},}\ }\href {\doibase 10.1063/1.446610} {\bibfield
  {journal} {\bibinfo  {journal} {The Journal of Chemical Physics}\ }\textbf
  {\bibinfo {volume} {80}},\ \bibinfo {pages} {5832--5838} (\bibinfo {year}
  {1984})}\BibitemShut {NoStop}%
\bibitem [{\citenamefont {Stone}\ and\ \citenamefont
  {Wales}(1986)}]{Stone1986}%
  \BibitemOpen
  \bibfield  {author} {\bibinfo {author} {\bibfnamefont {A.~J.}\ \bibnamefont
  {Stone}}\ and\ \bibinfo {author} {\bibfnamefont {D.~J.}\ \bibnamefont
  {Wales}},\ }\bibfield  {title} {\enquote {\bibinfo {title} {{Theoretical
  studies of icosahedral C60 and some related species}},}\ }\href {\doibase
  10.1016/0009-2614(86)80661-3} {\bibfield  {journal} {\bibinfo  {journal}
  {Chemical Physics Letters}\ }\textbf {\bibinfo {volume} {128}},\ \bibinfo
  {pages} {501--503} (\bibinfo {year} {1986})}\BibitemShut {NoStop}%
\bibitem [{\citenamefont {Ma}\ \emph {et~al.}(2009)\citenamefont {Ma},
  \citenamefont {Alf{\`{e}}}, \citenamefont {Michaelides},\ and\ \citenamefont
  {Wang}}]{Ma2009}%
  \BibitemOpen
  \bibfield  {author} {\bibinfo {author} {\bibfnamefont {J.}~\bibnamefont
  {Ma}}, \bibinfo {author} {\bibfnamefont {D.}~\bibnamefont {Alf{\`{e}}}},
  \bibinfo {author} {\bibfnamefont {A.}~\bibnamefont {Michaelides}}, \ and\
  \bibinfo {author} {\bibfnamefont {E.}~\bibnamefont {Wang}},\ }\bibfield
  {title} {\enquote {\bibinfo {title} {{Stone-Wales defects in graphene and
  other planar sp2-bonded materials}},}\ }\href {\doibase
  10.1103/PhysRevB.80.033407} {\bibfield  {journal} {\bibinfo  {journal}
  {Physical Review B}\ }\textbf {\bibinfo {volume} {80}},\ \bibinfo {pages}
  {033407} (\bibinfo {year} {2009})}\BibitemShut {NoStop}%
\bibitem [{\citenamefont {Zhang}\ \emph {et~al.}(2016)\citenamefont {Zhang},
  \citenamefont {Zhang}, \citenamefont {Ye},\ and\ \citenamefont
  {Zheng}}]{Zhang2016}%
  \BibitemOpen
  \bibfield  {author} {\bibinfo {author} {\bibfnamefont {J.}~\bibnamefont
  {Zhang}}, \bibinfo {author} {\bibfnamefont {H.}~\bibnamefont {Zhang}},
  \bibinfo {author} {\bibfnamefont {H.}~\bibnamefont {Ye}}, \ and\ \bibinfo
  {author} {\bibfnamefont {Y.}~\bibnamefont {Zheng}},\ }\bibfield  {title}
  {\enquote {\bibinfo {title} {{Free-end adaptive nudged elastic band method
  for locating transition states in minimum energy path calculation}},}\ }\href
  {\doibase 10.1063/1.4962019} {\bibfield  {journal} {\bibinfo  {journal} {The
  Journal of Chemical Physics}\ }\textbf {\bibinfo {volume} {145}},\ \bibinfo
  {pages} {094104} (\bibinfo {year} {2016})}\BibitemShut {NoStop}%
\end{thebibliography}%

\end{document}